\documentclass[aps,prd,amsmath,floats,floatfix,twocolumn,superscriptaddress,nofootinbib,noshowpacs]{revtex4-2}

\usepackage{amssymb}
\usepackage{amsmath}
\usepackage{physics}
\usepackage{latexsym}

\usepackage{verbatim}
\usepackage{mathrsfs}
\usepackage{amsfonts}

\usepackage{graphicx,subfigure}
\usepackage{epsfig}
\usepackage{pgfplots}

\usepackage{units}
\usepackage{overpic}
\usepackage[utf8]{inputenc}
\usepackage{rotating}
\usepackage{enumerate}

\usepackage{xfrac}
\usepackage{color}
\usepackage{multirow}

\newcommand\scalemath[2]{\scalebox{#1}{\mbox{\ensuremath{\displaystyle #2}}}}

\begin{document}
\title{On the radial linear stability of nonrelativistic $\ell$-boson stars}
\author{Armando A. Roque}
\affiliation{Instituto de F\'isica y Matem\'aticas,
Universidad Michoacana de San Nicol\'as de Hidalgo,
Edificio C-3, Ciudad Universitaria, 58040 Morelia, Michoac\'an, M\'exico}
\affiliation{Unidad Acad\'emica de F\'isica, Universidad Aut\'onoma de Zacatecas, 98060, M\'exico.}
\author{Emmanuel Ch\'avez Nambo}
\author{Olivier Sarbach}
\affiliation{Instituto de F\'isica y Matem\'aticas,
Universidad Michoacana de San Nicol\'as de Hidalgo,
Edificio C-3, Ciudad Universitaria, 58040 Morelia, Michoac\'an, M\'exico}
\date{\today}
\begin{abstract}
We study the linear stability of nonrelativistic $\ell$-boson stars, describing static, spherically symmetric configurations of the Schr\"odinger-Poisson system with multiple wave functions having the same value of the angular momentum $\ell$. In this work we restrict our analysis to time-dependent perturbations of the radial profiles of the $2\ell+1$ wave functions, keeping their angular dependency fixed. Based on a combination of analytic and numerical methods, we find that for each $\ell$, the ground state is linearly stable, whereas the $n$'th excited states possess $2n$ unstable (exponentially in time growing) modes. Our results also indicate that 
all excited states correspond to saddle points of the conserved energy functional of the theory.
\end{abstract}

\maketitle

\section{Introduction}

$\ell-$boson stars~\cite{Alcubierre:2018ahf} are exotic compact objects composed of $N=2\ell + 1$ self-gravitating complex massive scalar fields. They constitute a generalization of the standard ($\ell=0$) boson stars~\cite{PhysRev.172.1331, PhysRev.187.1767, 1992PhR...220..163J, Schunck:2003kk, Liebling:2012fv, Visinelli:2021uve} in which the internal symmetry group is extended from $U(1)$ to $U(N)$. This allows one to construct configurations in which each scalar field has the same harmonic time-dependency and carries angular momentum $\ell$; yet as a whole, the configuration is static and spherically symmetric~\cite{Olabarrieta:2007di, Alcubierre:2018ahf}. A recent interpretation of $\ell$-boson stars and more general configurations in semiclassical gravity, which only requires a single and real (quantum) scalar field, was recently presented in Ref.~\cite{Alcubierre:2022rgp}.

Similar to other compact objects (see e.g.,~\cite{Cardoso:2019rvt} for a review and~\cite{Barranco:2021auj, Roque:2021lvr} for more recent work in Horndeski theory), $\ell-$boson stars present a rich phenomenology~\cite{ Alcubierre:2021psa,Sanchis-Gual:2021edp, Jaramillo:2022zwg}. In particular, it has been shown that -- similarly to fluid stars -- these objects possess a ``stable branch", that is, configurations which are stable with respect to spherically symmetric linear~\cite{Gleiser:1988rq, Gleiser:1988ih, Alcubierre:2021mvs} and non-linear~\cite{Hawley:2000dt, Alcubierre:2019qnh} perturbations. Furthermore, full 3D numerical simulations~\cite{Jaramillo:2020rsv} of the Einstein-Klein-Gordon equations have found no indication of nonspherical growing perturbations for these configurations. However, the limited time scale of the simulations makes it difficult to reach a firm conclusion regarding the stability of these objects with respect to generic small perturbations.

In the Newtonian limit, $\ell$-boson stars are expected to reduce to solutions of the Schr\"odinger-Poisson (SP) system. For $\ell=0$ such solutions have been discussed in a variety of different physical contexts, including the Hartree-Fock theory of plasmas~\cite{Lieb1977}, the discussion of quantum state reduction by gravity~\cite{1998MPLA...13.2327B, Moroz:1998dh}, and the modeling of dark matter galactic halos~\cite{Schive:2014dra, Schive:2014hza, Marsh:2015wka, Gonzalez-Morales:2016yaf}. In particular, the following results have been established for the $\ell=0$ configurations. Lieb~\cite{Lieb1977} showed the existence of a unique stationary (that is, with the wave function having a time-harmonic dependency) ground state solution which is spherically symmetric and positive. The global in-time well-posedness of the Cauchy problem for the SP system has been shown in~\cite{Ginibre1980OnAC, Reinhard1994}. Further development led to the orbital stability of the ground state configurations~\cite{Cazenave1982}. The existence of excited spherical states, in which the radial profile of the wave function has any number $n$ of zeros, has been established in~\cite{1999Nonli..12..201T}. This leads to an infinite family of stationary, spherically symmetric solutions with negative energy eigenvalues which increase monotonically in $n$. For a recent review on the mathematical properties of solutions of the SP and related system, see~\cite{Moroz2016}. We also refer the reader to~\cite{KAVIAN2015942} for a recent existence result of the SP system with infinitely many states on a bounded domain. Regarding the generalization to $\ell$-boson stars with $\ell\neq 0$, in~\cite{nambo21} the existence of an infinite family of stationary, spherically symmetric solutions to the SP system with $n$ nodes in the wave functions radial profile has been proven for each $n,\ell = 0,1,2,3,\ldots$. For numerical examples of these configurations, see~\cite{jaramillo19, nambo19, nambo21}.

Similar to their relativistic counterparts, the realization of these objects in Nature demands that they are stable with respect to sufficiently small dynamical perturbations.\footnote{Or, in case they are unstable, have a sufficiently large lifetime.} In the $\ell=0$ case, a stability analysis has been performed in~\cite{2002math.ph...8045H}, based on a combination of analytic and numerical methods. Their results indicate that the ground state configuration is stable, whereas each excited state is unstable, having $n$ quadruples of modes characterized by a complex eigenvalue. For further numerical studies regarding the nonlinear stability of the $\ell=0$ ground base configurations, see Refs.~\cite{Guzman:2004wj, Bernal:2006it}. For a generalization to multi-state solutions (having $\ell=0$ but different $n$'s) see~\cite{Urena-Lopez:2010zva}. 3D numerical evolutions of the multistate SP system which analyze the stability of multi-$\ell$ multi-state configurations (i.e., solutions containing wave functions with multiple values of $\ell$ and $n$) and other configurations which are axially symmetric have recently been performed in~\cite{Guzman:2019gqc}. In particular, it is claimed in that work that the nonrelativistic ground state $\ell$-boson stars with $\ell=1$ are stable.\footnote{Note that the definition of the quantum number $n$ in~\cite{Guzman:2019gqc} differs from our definition; it is such that the node number of the radial wave function is equal to $n-1-\ell$.} However, as far as we are aware, no linear stability analysis for the $\ell\neq 0$ configurations has been carried out so far.

The goal of this article is to provide a systematic study of the nonrelativistic $\ell$-boson stars' main properties and to analyze their mode stability with respect to spherically symmetric linear perturbations. To this purpose, we start in Sec.~\ref{SecII} with a description of the theoretical framework underlying the construction of these objects and their linear perturbations, starting with the $N$-particle SP system. Of particular relevance for this work is the identification of a time-conserved energy functional describing the total energy of the system. This functional is an extension to $N$ particles of the well-known functional used in~\cite{Lieb1977}, whose global minimum describes the $\ell=0$ ground state. More generally, as we show, the nonrelativistic $\ell$-boson stars correspond to critical points of this functional. The reduction to one-particle states is presented in subsection~\ref{IIA}, whereas the spherically symmetric system, in which the wave functions are assumed to have the same radial profile with particular angular dependencies, is derived in~\ref{IIB}. The stationary and linearized equations are presented in subsections~\ref{IIC} and \ref{IID}, respectively. Analytic properties of the system, including its rescaling freedom, a zero mode solution, and a fourfold symmetry between the mode solutions of the linearized equations, are studied in subsection~\ref{IIE}. Also in this subsection, we show that the linearized system can be reduced to a single equation involving a linear operator $\hat{Q}$ which is related to the second variation of the energy functional.

Section~\ref{SecIII} presents our numerical implementation and the results for the nonrelativistic $\ell$-boson stars. In subsection~\ref{IIIA} we rewrite the stationary system in a more suitable form for the numerical calculations and analyze the regularity conditions at the center and the asymptotic behavior of the solutions at infinity. Further, we explain our shooting method used for the computation of the wave function and their energy eigenvalues. Next, in subsection~\ref{IIIB} we exhibit the numerical results for $n=0,1, 2, 3$ and $\ell = 0, 1, 2, 3, 4, 5$. In particular, we show the radial profiles of the wave function and the gravitational potential and provide a table for the energy eigenvalues. Also, we study the total energy of the system, identifying energetically allowed transitions between configurations with the same total number $N$ of fields. This analysis indicates that the ground state configuration with zero angular momentum corresponds to the global minimum of the conserved energy functional. In fact, this property can be established from the results in~\cite{Lieb1977}.

Section~\ref{SecIV} is devoted to the numerical study of the linearized system. Our method is a straightforward generalization of the procedure used in~\cite{2002math.ph...8045H} to arbitrary values of $\ell$, in which the linearized equations are reduced to an eigenvalue problem which is solved by spectral methods. In subsection~\ref{IVA} these equations are rewritten in a more appropriate form, and the physically relevant boundary conditions at the origin and the asymptotic region, which are then used to provide boundary conditions at a finite outer boundary, are derived. Further, we describe our procedure for solving the eigenvalue problem via a pseudo-spectral collocation method with Chebyshev points~\cite{trefethen2000spectral, boyd2013chebyshev}. In subsection~\ref{IVB} we exhibit the numerical linear stability results, firstly for configurations in the ground state and next for the excited states. In both cases, we show the spectrum of eigenvalues and the profiles of the associated eigenfunctions for some representative examples, and we comment on the lifetime of the unstable configurations. Our results indicate that for all $\ell\geq 0$ the ground state solution is stable and corresponds to a minimum of the (spherically symmetric reduced) energy functional, whereas all excited states are unstable and correspond to saddle points. Further, our results suggest that the ground states possess only purely oscillatory modes.

Conclusions and open questions are provided in section~\ref{SecV}. Technical results, which include a Lagrangian formulation of the SP system, the computation of the first and second variations of the conserved energy functional, the numerical determination of the energy eigenvalues and a validation of our numerical spectral code, are included in appendixes.
 
\section{Theoretical setup}
\label{SecII}

Consider a non-relativistic system consisting of $N$ identical particles of mass $\mu$ whose only interaction is through the gravitational potential $\mathbf{U}(t,\vec{x})$ generated by them. Such a system is described by the $N$-particle Schr\"odinger-Poisson (or gravitational Schr\"odinger) system~\cite{Diosi:1984wuz, Jones:1995yz, Jones:1995wb}
\begin{align}
\scalemath{0.98}{
	i\hbar\frac{\partial \Psi(t,X)}{\partial t}= \sum_{i=1}^{N}\left( -\frac{\hbar^2}{2\mu}\laplacian_{\vec{x}_i}+\mu \mathbf{U}(t, \vec{x}_i)\right) \Psi(t,X),\label{eqSID}
 }
\end{align}
where $\hbar$ denotes the reduced Planck constant, $\Psi(t,X)$ is the wave function with $X = (\vec{x}_1,\vec{x}_2,\dots, \vec{x}_N)$ the $3N$-vector parameterizing the configuration space. Here, $\nabla^2_{\vec{x}_i}$ refers to the 3D Laplace operator with respect to the variable $\vec{x}_i$, and the gravitational potential $\mathbf{U}(t,\vec{x})$ generated by the $N$ particles is determined by the Poisson equation
\begin{align}
\scalemath{0.93}{
    \laplacian{\mathbf{U}}(t, \vec{x})=4\pi G \mu\sum_{i=1}^N\int\abs{\Psi(t,X)}^2\delta^{(3)}(\vec{x}-\vec{x}_i)d^{3N}X,\label{eqSU}
    }
\end{align}
with the requirement that ${\mathbf{U}}(t, \vec{x})\to 0$ for $|\vec{x}|\to \infty$. Using Green's function of the Laplace operator, one can represent $\mathbf{U}$ as
\begin{align}
    \mathbf{U}(t, \vec{x})&=-G\mu \sum_{j=1}^N\int\frac{\abs{\Psi(t, Y)}^2}{\abs{\vec{x}-\vec{y}_{j}}}d^{3N}Y,
\label{eqPotential}
\end{align}
where the integral is performed over the $3N$-vector $Y=(\vec{y}_1,\dots,\vec{y}_N)$ and $G$ refers to Newton's constant.

The evolution described by the nonlinear system~(\ref{eqSID}, \ref{eqPotential}) is unitary, i.e., the $L^2$-norm of the wave function $\Psi$ is preserved in time. Additionally, it is straightforward to verify that the functional (cf.~\cite{Lieb1977, Cazenave1982, Diosi:1984wuz, Jones:1995yz, Jones:1995wb})
\begin{align}
\mathcal{E}[u] &= \sum_{i=1}^N \frac{\hbar^2}{2\mu}\int \abs{\nabla_{\vec{x}_i} u(X)}^2 d^{3N}X\nonumber\\
    &-\frac{G\mu^2}{2}\sum_{i, j=1}^N \int\int\frac{\abs{u(X)}^2 \abs{u(Y)}^2}{\abs{\vec{x}_i-\vec{y}_j}} d^{3N}X d^{3N}Y,\label{ecFuncGen}
\end{align}
is conserved in time, that is $\mathcal{E}[\Psi(t,\cdot)]$ is independent of $t$ for any solution $\Psi(t,X)$ of Eqs.~(\ref{eqSID}, \ref{eqPotential}) for which $|\mathcal{E}[\Psi(0,\cdot)]| < \infty$. As discussed in the next two sections, its second variation will be very useful to understand the stability properties of the $\ell$-boson stars.

\subsection{Reduction to one-particle states}\label{IIA}

From now on, we focus on the particular case in which the particles are indistinguishable and spinless. Furthermore, we assume that these are uncorrelated\footnote{This ansatz is valid because our model assumes that the particles do not interact directly between themselves; they only interact through the common ``mean field" Newtonian potential $\mathbf{U}$ they generate. For more details on this separability property, see Refs.~\cite{Bialynicki-Birula:1976tja, Diosi:1984wuz}. For the stationary case, the ansatz~(\ref{ansatz1}) is equivalent to the Hartree approximation~\cite{1977CMaPh..53..185L}.}, such that the $N$-particle wave function is a (symmetrized) product of single-particle states. Specifically, we consider an orthonormal set of wave functions $\psi_j$ in the one-particle Hilbert space $L^2(\mathbb{R}^3)$, such that $(\psi_j,\psi_k)=\delta_{jk}$. Assuming that there are $N_j$ particles in the state $\psi_j$, the $N$-particle wave function can be written as
\begin{equation}
\Psi = \sqrt{\frac{N!}{N_1! N_2!\cdots N_J!}}
\hat{S}\left(\psi_1^{N_1}\otimes \psi_2^{N_2}\otimes\ldots \otimes \psi_J^{N_J} \right),\label{ansatz1}
\end{equation}
where $\sum_{j=1}^J N_j = N$. Here, $\hat{S}=\sum_{\pi\in \sigma(N)} P_\pi/N!$ denotes the symmetrization operator (with $\sigma(N)$ referring to the permutation group of $N$ elements and $P_\pi$ to the permutation operator). Introducing the ansatz~(\ref{ansatz1}) into Eqs.~(\ref{eqSID},\ref{eqSU}), one finds that the one-particle wave functions $\psi_j$ satisfy the system
\begin{subequations}\label{eqT}
    \begin{align}
    	i\hbar\frac{\partial \psi_{j}(t, \vec{x})}{\partial t}&= \left( -\frac{\hbar^2}{2\mu}\laplacian +\mu \mathbf{U}(t, \vec{x})\right) \psi_{j}(t, \vec{x}),\label{eqT1}\\
    	\laplacian{\mathbf{U}}(t, \vec{x})&=4\pi G \mu\sum_{j=1}^J N_j \abs{\psi_{j}(t, \vec{x})}^2.\label{eqT2}
    \end{align}
\end{subequations}
It is not difficult to prove that the evolution preserves each scalar product $(\psi_j,\psi_k)$, such that it is sufficient to impose the orthonormality condition $(\psi_j,\psi_k) = \delta_{jk}$ at the initial time $t=0$. Furthermore, the functional $\mathcal{E}[u]$ reduces to
\begin{align}
\mathcal{E}&[u] = \sum_{j=1}^J N_j \frac{\hbar^2}{2\mu}\int \abs{\nabla u_j(\vec{x})}^2 d^3 x\nonumber\\
    &-\frac{G\mu^2}{2}\sum_{j, k=1}^J N_j N_k \int\int\frac{\abs{u_j(\vec{x})}^2 \abs{u_k(\vec{y})}^2}{\abs{\vec{x}-\vec{y}}} d^3x d^3y,\label{ecFuncGenReduced}
\end{align}
where the relation between $u$ and $u_j$ is the same as the one between $\Psi$ and $\psi_j$ in Eq.~(\ref{ansatz1}).

\subsection{Spherically symmetric system}\label{IIB}

The standard solutions of the SP system correspond to the particular case $N = J = 1$ in which there is only one wave function. However, allowing the presence of an arbitrary number $N$ of particles yields a much richer model, even when restricted to spherically symmetric configurations.

Like their relativistic counterparts, Newtonian $\ell$-boson stars are obtained by considering $N = J = 2\ell+1$ particles in a spherically symmetric static potential with associated wave functions of the form
\begin{equation}
\psi_{j}(t, \vec{x}):=f_{\ell}(t, r)Y^{\ell m}(\vartheta, \varphi).
\label{ansatzAng}
\end{equation}
Here, $Y^{\ell m}$ denote the standard spherical harmonics, and $f_\ell$ a function describing the radial profile which has $n$ nodes in the interval $0 < r < \infty$ and is identical for all states. In other words, $\ell$-boson stars are characterized by the quantum numbers $(n,\ell,m)$, where $n$ and $\ell$ are fixed and $m$ varies over $-\ell,-\ell+1,\ldots,\ell$. Accordingly, the relation between the index $j$ and $(n,\ell,m)$ in Eq.~(\ref{ansatzAng}) is given by $j = m+\ell+1$. A simple generalization of $\ell$-boson stars consists in occupying each state $\psi_j$ with $K$ particles instead of just one, such that $N = K J = K(2\ell+1)$. 

Introducing the ansatz~(\ref{ansatzAng}) into the system~(\ref{eqT}) and taking into account the identities $\nabla^2 Y^{\ell m}=-\ell(\ell+1)Y^{\ell m}/r^2$ and $\sum_{m=-\ell}^\ell\abs{Y^{\ell m}}^2=(2\ell+1)/(4\pi)$, one obtains
\begin{subequations}\label{ellSis}
\begin{align}
    i\hbar\frac{\partial f_{\ell}(t, r)}{\partial t}=& \Bigg[ \frac{\hbar^2}{2\mu}\left(-\laplacian_s+\frac{\ell(\ell+1)}{r^2}\right)+\mu \mathbf{U}(t, r)\Bigg] f_{\ell}(t, r),\\
	\laplacian_s{\mathbf{U}(t, r)}=&  (2\ell+1) \mu G K \abs{f_{\ell}(t, r)}^2,
\end{align}
\end{subequations}
where here and in the following, $\laplacian_s := \frac{1}{r^2}\frac{\partial}{\partial r}\left(r^2\frac{\partial}{\partial r}\right)$ denotes the radial part of the Laplacian. Note that the effect of including the occupation number $K$ is formally equivalent to rescaling Newton's constant $G$.

For the following, it is convenient to rewrite this system in terms of dimensionless quantities. To this purpose, first note that $G,\hbar,\mu$ give rise to a characteristic distance and length defined by
\begin{subequations}
\begin{align}
d_c &:= \frac{\hbar^2}{2G\mu^3}= 1.78048\times 10^{22}\left(\frac{m_p}{\mu}\right)^{3}\; \text{m},\\
t_c &:= \frac{\hbar^3}{2G^2\mu^5}=1.0056\times 10^{52}\left(\frac{m_p}{\mu}\right)^{5} \; s,
\end{align}
\end{subequations}
where, as reference, we have specified the numerical values resulting from the proton mass $m_p = 1.67262 \times 10^{-27} \text{kg}$. Next, we introduce the transformation
\begin{align}\label{AdVar}
    \begin{array}{ll}
        t=t_c \bar{t}/(K\Lambda)^{2}, & r= d_c\bar{r}/(K\Lambda),\\[0.3cm]
        f_{\ell} = K^{3/2} \Lambda^2\bar{f}_{\ell}/\sqrt{(2\ell+1)d_c^3}, & \mathbf{U}=2 v_c^2 (K\Lambda)^{2}\bar{U},
    \end{array}
\end{align}	
with $v_c := d_c/t_c$ a characteristic velocity. Here, the bar refers to dimensionless quantities and $\Lambda$ is an arbitrary positive dimensionless scale factor. In order to simplify the notation, in what follows we shall omit the bars and denote dimensionfull quantities with the superscript $phys$ whenever necessary.

Performing the transformation described in Eq.~(\ref{AdVar}), the system~(\ref{ellSis}) reduces to
\begin{subequations}\label{ellSis1}
\begin{align}
	i\frac{\partial f_{\ell}(t, r)}{\partial t}&= \left[-\laplacian_s+\frac{\ell(\ell+1)}{r^2}+ U(t, r)\right] f_{\ell}(t, r),\\
	\laplacian_s{U(t, r)}&= \abs{f_{\ell}(t, r)}^2.
\end{align}
\end{subequations}
The normalization condition $(\psi_j,\psi_j) = 1$ is satisfied provided that
\begin{equation}\label{EqNorma}
   1= \int\abs{\psi_j(t, \vec{x})}^2 d^3 x = \frac{\Lambda}{2\ell+1}\int_{0}^{\infty}\abs{f_{\ell}(t, r)}^2 r^2 dr.
\end{equation}
Equivalently, the system~(\ref{ellSis1}) can be written as the single nonlinear equation
\begin{equation}
	i\frac{\partial f_{\ell}(t, r)}{\partial t}= \hat{\mathcal{H}_\ell} f_{\ell}(t, r),\label{eqSIDl}
\end{equation}
with the integro-differential operator
\begin{equation}
	\hat{\mathcal{H}}_\ell := \left[ -\nabla_s^2+\frac{\ell(\ell+1)}{r^2}
+\triangle^{-1}_{s}(\abs{f_{\ell}(t, \cdot)}^2)
\right].
\label{HOp1}
\end{equation}
Here, $\triangle_s^{-1}$ denotes the inverse of $\nabla_s^2$, defined by
\begin{equation}
\triangle^{-1}_s(A) (r) :=-\int_0^\infty \frac{A(\tilde{r})}{r_>} \tilde{r}^2 d\tilde{r},\label{LaplInve}
\end{equation}
when acting on an arbitrary function $A$ depending only on the radius $r$, where we have set $r_{>}:=\max \left\lbrace r, \tilde{r} \right\rbrace$.

For the particular subset of solutions of the form~(\ref{ansatzAng}), the conserved energy functional~(\ref{ecFuncGenReduced}) reduces to
\begin{align}
\mathcal{E}^{phys} = 2\mu v_c^2(\Lambda K)^3\mathcal{E}_{\ell}[f_\ell],
\label{EnerRel}
\end{align}
where the dimensionless functional $\mathcal{E}_{\ell}$ is given by
\begin{align}
    \mathcal{E}_{\ell}[f] &= \int_0^\infty \left[ \abs{\partial_r f(r)}^2+\frac{\ell(\ell+1)\abs{f(r)}^2}{r^2}\right] r^2 dr\nonumber\\
    &-\frac{1}{2} \int_0^\infty\int_0^\infty\frac{\abs{f(r)}^2 \abs{f(\tilde{r}))}^2}{r_{>}} r^2 \tilde{r}^2 dr d\tilde{r}.
\label{ecFuncellBS}
\end{align}
Note that the scale factor $\Lambda$ offers the possibility to solve the system~(\ref{ellSis1}) or Eq.~(\ref{eqSIDl}) without taking into account the normalization condition~(\ref{EqNorma}) in a first step. Equation~(\ref{EqNorma}) can be enforced in a second step by adjusting the value of $\Lambda$.

\subsection{The stationary equations}\label{IIC}

The nonrelativistic $\ell$-boson stars are obtained as solutions of Eq.~(\ref{eqSIDl}) with the time-harmonic ansatz
\begin{equation}\label{Hansatz}
f_{\ell}(t, r) 
 = e^{-i E_{\ell} t}\sigma^{(0)}_{\ell}(r),
\end{equation}
where $\sigma^{(0)}_\ell$ is a real-valued radial function and $E_\ell$ is determined by the nonlinear eigenvalue problem
\begin{equation}
\label{SecOrd02lB}
 \hat{\mathcal{H}_\ell}^{(0)}\sigma_\ell^{(0)}=E_\ell\sigma_\ell^{(0)},
\end{equation}
with 
\begin{align}
 \hat{\mathcal{H}}_\ell^{(0)}&:= \left[ -\nabla_s^2+\frac{\ell(\ell+1)}{r^2}+\triangle^{-1}_{s}\left(\abs{\sigma_\ell^{(0)}}^2\right)\right].\label{Eq:Hell0}
\end{align}
The eigenvalue $E_{\ell}$ represents the energy of each state $(n,\ell,m)$. In physical units, this energy value reads
\begin{equation}
E^{phys}_{\ell}=2\mu v_c^2(K \Lambda)^2 E_{\ell}.
\end{equation}
The existence of a normalizable solution of Eq.~(\ref{SecOrd02lB}) for each value of $n$ and $\ell$ has been established in~\cite{nambo21}. In the next section, we implement a numerical method to solve Eq.~(\ref{SecOrd02lB}) that generalizes the procedure presented in~\cite{Moroz:1998dh} to arbitrary $\ell$. As shown in Appendix~\ref{App:LagrangeFormulation} the conserved energy functional for any stationary solutions of the form~(\ref{Hansatz}) takes the value
\begin{equation}
\mathcal{E}^{phys} = K\frac{2\ell+1}{3} E_\ell^{phys}.
\label{Eq:FunctionalValueLBS}
\end{equation}
For $\ell=0$ and $K=1$ this reduces to the well-known relation $\mathcal{E}^{phys} = E_0^{phys}/3$ presented in~\cite{TOD2001173} for the standard boson star solutions (note that the functional $I$ in~\cite{TOD2001173} satisfies $2I = \mathcal{E}^{phys}$). Equation~(\ref{Eq:FunctionalValueLBS}) will turn out to be useful when comparing the ground state energies of different families of $\ell$-boson stars with each other.

\subsection{The linearized equations}\label{IID}

Next, we proceed to linearize the integro-differential equation~(\ref{eqSIDl}) about a stationary solution. To this purpose, we assume an expansion of $f_{\ell}$ in terms of a small parameter $0 < \epsilon\ll 1$ of the form
\begin{equation}
f_{\ell}(t, r) = e^{-i E_\ell t}\left[ \sigma^{(0)}_{\ell}(r) 
 + \epsilon\sigma_{\ell}(t, r) + {\mathcal O}(\epsilon^2) \right],
\label{eq:ansatzPert}
\end{equation}
where $(E_\ell, \sigma^{(0)}_{\ell})$ is a solution of the nonlinear eigenvalue problem~(\ref{SecOrd02lB}) and $\sigma_{\ell}$ is a complex-valued function depending on $(t,r)$ which describes the linear perturbation. Following~\cite{2002math.ph...8045H} we separate the temporal and radial parts of this function by means of the following ansatz (see also Sec.~5.2 in~\cite{10.5555/1941970} for details):
\begin{equation}
\sigma_{\ell}(t,r) = \left[A(r)+B(r)\right]e^{\lambda t}+\left[A(r)-B(r)\right]^{*}e^{\lambda^* t},
\label{Ecpert}
\end{equation}
where $A$ and $B$ are complex-valued functions depending only on $r$, $\lambda$ is a complex constant and the superscript ${}^*$ denotes complex conjugation. A linear instability is signaled by the presence of a solution with a positive real part of $\lambda$. Introducing Eqs.~(\ref{eq:ansatzPert}, \ref{Ecpert}) into Eq.~(\ref{eqSIDl}) one obtains, to linear order in $\epsilon$ and after setting the coefficients in front of $e^{\lambda^* t}$ and $e^{\lambda t}$ to zero,
\begin{subequations}\label{SecOrd12lB}
 	\begin{eqnarray}
        i\lambda A&=&\left(\hat{\mathcal{H}_\ell}^{(0)}-E_\ell\right)B, \label{SecOrd12blB}\\
 		i\lambda B&=&\left(\hat{\mathcal{H}_\ell}^{(0)}-E_\ell\right) A+2\sigma_\ell^{(0)}\triangle^{-1}_{s}\left[\sigma_\ell^{(0)}A\right].\label{SecOrd12alB}
 	\end{eqnarray}
 \end{subequations}
This system constitutes a linear eigenvalue problem for the eigenvalue $\lambda$. In the next subsection, we derive some basic properties satisfied by the solutions of Eqs.~(\ref{eqSIDl}) and~(\ref{SecOrd12lB}).

\subsection{Basic properties of the solutions}\label{IIE}

As follows from Eq.~(\ref{AdVar}) the system~(\ref{ellSis1}) has the following rescaling freedom: given a solution $(f_{\ell}, U)$, then
\begin{subequations}\label{reescaling}
\begin{align}
f^\Lambda_{\ell}(t, r)&=\Lambda^2 f_{\ell}(\Lambda^{2} t, \Lambda r),\\
U^\Lambda(t, r)&=\Lambda^2 U(\Lambda^{2} t, \Lambda r),
\end{align}
\end{subequations}
is also a solution of the system~(\ref{ellSis1}). As pointed out previously, this freedom offers the possibility to look for a solution $(k_{\ell},\Phi)$ of Eq.~(\ref{ellSis1}) whose normalization is arbitrary but finite. The correct normalization condition~(\ref{EqNorma}) can be enforced a posteriori by means of the transformation $(f_\ell,U) = (k_\ell^\Lambda,\Phi^\Lambda)$ with
\begin{equation}
\Lambda :=\frac{2\ell+1}{\int_0^\infty\abs{k_{\ell}(t, r)}^2 r^2 dr}. \label{LambConst}
\end{equation}
In the stationary case, the rescaling induces the transformation $E^\Lambda_\ell = \Lambda^2 E_\ell$ for the energy eigenvalues in Eq.~(\ref{Hansatz}).

Next, we discuss a few properties of the system of linearized equations~(\ref{SecOrd12lB}). First, note the existence of the zero mode solution $(A,B) = (0,\beta\sigma_{\ell}^{(0)})$ with eigenvalue $\lambda=0$ and an arbitrary complex constant $\beta$. For $\beta = i$ this solution corresponds to an infinitesimal rotation in the phase of the unperturbed wave function, whereas $\sigma_\ell(t,r)$ is identically zero if $\beta$ is real. Second, it is simple to prove that any solution $(\lambda, A, B)$ of the linearized system~(\ref{SecOrd12lB}) gives rise to the three other solutions:
\begin{align}
\begin{array}{lcr}
(-\lambda,A,-B),
& (\lambda^*,A^*,-B^*),
& (-\lambda^*,A^*,B^*).
\label{Eq:lambda4}
\end{array}
\end{align}
Therefore, the eigenvalues come in pairs $(\lambda,-\lambda)$ if $\lambda$ is real or purely imaginary and in quadruples $\{\lambda, -\lambda, \lambda^{*}, -\lambda^{*}\}$ otherwise, and their corresponding eigenfunctions $(A, B)$ are related to each other (up to a global factor) according to Eq.~(\ref{Eq:lambda4}). Therefore, linear stability requires that the real part of each eigenvalue $\lambda$ is zero.

Third, we note the following properties. Multiplying Eq.~(\ref{SecOrd12blB}) with $r^2 A^*$ and Eq.~(\ref{SecOrd12alB}) with $r^2 B^*$ and integrating yields
\begin{subequations}
\begin{align}
 i\lambda(B,A)_{L^2} =& \left( B, \left[\hat{\mathcal{H}_\ell}^{(0)}-E_\ell\right] B \right)_{L^2},
 \label{Eq:ilBA}\\
 i\lambda(A,B)_{L^2} =& \left( A, \left[\hat{\mathcal{H}_\ell}^{(0)}-E_\ell\right] A \right)_{L^2}\nonumber\\
 &+ 2\left( \sigma_\ell^{(0)} A,\triangle^{-1}_{s}[\sigma_\ell^{(0)} A]
 \right)_{L^2},
 \label{Eq:ilAB}
\end{align}
\end{subequations}
where from now on, $(\cdot,\cdot)_{L^2}$ refers to the standard scalar product in the Hilbert space $L^2 = L^2(\mathbb{R}_+,r^2 dr)$, that is,
\begin{equation}
(A_1,A_2)_{L^2}:=\int_0^\infty A_1(r)^* A_2(r) r^2 dr,\qquad A_i\in L^2. \label{innerProd}
\end{equation}
Since the operator $\hat{\mathcal{H}_\ell}^{(0)}-E_\ell$ is self-adjoint in $L^2$, it follows that the right-hand side of Eq.~(\ref{Eq:ilBA}) is real. Likewise, the right-hand side of Eq.~(\ref{Eq:ilAB}) is real.\footnote{This follows again from the self-adjointness of $\hat{\mathcal{H}_\ell}^{(0)}-E_\ell$ and the identity
\begin{align}
\left(\sigma_\ell^{(0)}A, \triangle^{-1}_{s}\left [\sigma_\ell^{(0)}A\right]\right)_{L^2}
= -\left\| \nabla_s\bigg(\triangle^{-1}_{s}\left[\sigma_\ell^{(0)}A\right]\bigg) \right\|_{L^2}^2,
\nonumber
\end{align}
which shows that the second term on the right-hand side of Eq.~(\ref{Eq:ilAB}) is real and negative.}
Therefore, it follows that
\begin{equation}
\lambda^2|(A,B)_{L^2}|^2\in\mathbb{R},
\end{equation}
which implies that either $\lambda^2$ is real or $(A,B)_{L^2} = 0$. This generalizes the corresponding result in Ref.~\cite{2002math.ph...8045H} to arbitrary values of $\ell$.

Fourth, we note that the right-hand side of Eq.~(\ref{Eq:ilAB}) can also be written in terms of the second variation of the energy functional $\mathcal{E}_\ell$ (see Appendix~\ref{ApendVf} for a derivation):
\begin{equation}
i\lambda(A,B)_{L^2} 
 = \delta^2\mathcal{E}_{\ell}[A_R] 
 + \delta^2\mathcal{E}_{\ell}[A_I],
\label{Eq:ilABBis}
\end{equation}
where here and in the following, the subindices $R$ and $I$ refer to the real and imaginary parts of the field. Equation~(\ref{Eq:ilABBis}) will play an important role since it provides a direct relation between the eigenvalue and eigenfunctions of the linearized equations and the second variation of the energy functional.

Fifth, we note that the system~(\ref{SecOrd12lB}) can be reduced to the single equation
\begin{equation}
\hat{Q}A=-\lambda^2 A, \label{OpFO}
\end{equation}
with the fourth-order operator $\hat{Q}$ defined as
\begin{align}
\scalemath{0.95}{
\hat{Q}:=\left(\hat{\mathcal{H}_\ell}^{(0)}-E_\ell\right)^2+2\left(\hat{\mathcal{H}_\ell}^{(0)}-E_\ell\right)\sigma_\ell^{(0)}\triangle^{-1}_{s}\sigma_\ell^{(0)}.\label{EqOpe}
}
\end{align}
Note that the image of $\hat{Q}$ is orthogonal to $\sigma_\ell^{(0)}$. Indeed, using once again the self-adjointness of $\hat{\mathcal{H}_\ell}^{(0)}-E_\ell$ and the fact that $\sigma_\ell^{(0)}$ lies in its kernel one finds $(\sigma_\ell^{(0)}, \hat{Q}A)_{L^2}= 0$ for all $A$. Therefore, we may restrict the domain of $\hat{Q}$ to the subspace $Z$ consisting of the orthogonal complement of the background field $\sigma_\ell^{(0)}$ in $L^2$ and consider $\hat{Q}$ as an operator in the Hilbert space $Z$. Since $\hat{\mathcal{H}_\ell}^{(0)}-E_\ell$ is invertible on this subspace, we may equip $Z$ with the new inner product
\begin{align}
\braket{A_1}{A_2}:= \left( A_1,\left[\hat{\mathcal{H}_\ell}^{(0)}-E_\ell\right]^{-1}A_2 \right)_{L^2},\; A_i\in Z, \label{Eqiner}
\end{align}
which is such that
\begin{equation}
\braket{A_1}{\hat{Q}A_2} = \braket{\hat{Q}A_1}{A_2},
\label{Eq:QA12}
\end{equation}
for all $A_1,A_2$ lying in the domain of $\hat{Q}$. Therefore, $\hat{Q}$ is symmetric with respect to this new product. However, note that although this product is bounded and (anti-) linear and its (first) second argument, it is not always positive definite. If $\sigma_\ell^{(0)}$ is the ground state solution, such that $E_\ell$ is the minimum eigenvalue of $\hat{\mathcal{H}_\ell}^{(0)}$, then $\braket{\cdot}{\cdot}$ is positive definite on $Z$ and it follows from Eq.~(\ref{Eq:QA12}) that $\hat{Q}$ is (formally) self-adjoint, implying, in particular, that its eigenvalues $-\lambda^2$ are real. On the other hand, if $\sigma_\ell^{(0)}$ is an excited state with $n$ nodes, it follows from the nodal theorem (see e.g.,~\cite{bS05}) that $\hat{\mathcal{H}_\ell}^{(0)}$ possesses precisely $n$ eigenvalues smaller than $E_\ell$. Hence, the inner product $\braket{\cdot}{\cdot}$ has $n$ independent directions with negative norm, and in this case, there is no reason to expect that the eigenvalues of $\hat{Q}$ are real. It follows from Eqs.~(\ref{SecOrd12blB},~\ref{Eq:ilABBis},~\ref{OpFO}) that
\begin{equation}
-\lambda^2\braket{A}{A} = \braket{A}{\hat{Q}A} = \delta^2\mathcal{E}_{\ell}[A_R] 
 + \delta^2\mathcal{E}_{\ell}[A_I].
\label{Eq:FundamentalRelation}
\end{equation}

Based on these observations, we arrive at the following conclusions. For the ground state solutions with $n=0$, $\lambda^2$ is real (which implies that $\lambda$ itself is either real or purely imaginary) and the inner product $\braket{\cdot}{\cdot}$ is positive definite. Furthermore, since the operator $\hat{Q}$ is real, one can assume that any eigenfunction $A = A_R$ is real as well. In this case, Eq.~(\ref{Eq:FundamentalRelation}) implies that the signs of $-\lambda^2$ and $\delta^2\mathcal{E}_\ell[A]$ coincide with each other. Consequently, a purely imaginary point spectrum implies that the background solution is linearly stable and represents a local minimum of $\mathcal{E}_\ell$ (at least with respect to the space spanned by the eigenfunctions). However, the presence of a nonzero pure real eigenvalue would imply that the solution $\sigma_\ell^{(0)}$ is linearly unstable and correspondingly, there would exist a direction for which $\delta^2\mathcal{E}_\ell[A]$ is negative, meaning that $\sigma_\ell^{(0)}$ could not be a minimum of $\mathcal{E}_{\ell}$.

For the excited states (i.e., those with $n>0$ nodes), $\lambda^2$ does not need to be real, as commented above. To make further progress, we note that Eqs.~(\ref{SecOrd12lB}) and Eq.~(\ref{Ap1Eq9}) in Appendix~\ref{ApendVf} imply
\begin{subequations}\label{EqB3}
\begin{align}
\delta^2\mathcal{E}_{\ell}[A_R]
 &=-\lambda_R \left(A_R, B_I\right)_{L^2}-\lambda_I\left(A_R, B_R\right)_{L^2},  \label{EqB3.1} \\
\delta^2\mathcal{E}_{\ell}[A_I]
 &=\lambda_R\left(A_I, B_R\right)_{L^2}-\lambda_I\left(A_I, B_I\right)_{L^2}.  \label{EqB3.2} 
\end{align}
\end{subequations}
Taking into account the symmetries~(\ref{Eq:lambda4}) we have the following possibilities for the eigenvalue $\lambda$:
\begin{itemize}
\item[(i)] $\lambda_R=0, \lambda_I=0$: This is the zero mode solution we have already discussed above.
\item[(ii)] $\lambda_R > 0, \lambda_I = 0$: In this case, $\lambda$ is real and from Eq.~(\ref{SecOrd12lB}) we can assume that $A = A_R$ and $B = i B_I$. It follows from Eq.~(\ref{EqB3.1}) that $\delta^2\mathcal{E}_{\ell}[A_R] = -\lambda\left(A_R, B_I\right)_{L^2}$, such that the sign of the second variation of $\mathcal{E}_\ell$ is opposite to the sign of the product $\left(A_R, B_I\right)_{L^2}$.
\item[(iii)] $\lambda_R = 0, \lambda_I > 0$. In this case, $\lambda$ is purely imaginary and we can assume that both $A$ and $B$ are real. It follows from Eq.~(\ref{EqB3.1}) that the sign of $\delta^2\mathcal{E}_{\ell}[A_R]$ is opposite to the sign of $\left(A_R, B_R\right)_{L^2}$.
\item[(iv)] $\lambda_R > 0, \lambda_I > 0$. In this case, $\lambda^2\neq \mathbb{R}$, and from the previous points it follows that $(A,B)_{L^2} = 0$. In this case, it follows from Eq.~(\ref{Eq:ilABBis}) that
\begin{equation}
\delta^2\mathcal{E}_{\ell}[A_R] + \delta^2\mathcal{E}_{\ell}[A_I] = 0,
\end{equation}
such that the condition $\delta^2\mathcal{E}_{\ell}[A_R]\neq 0$ (which can be verified using Eq.~(\ref{EqB3.1})) implies that the background solution $\sigma_\ell^{(0)}$ corresponds to a critical saddle point of the energy functional. 
\end{itemize}
The first three possibilities also apply to the ground state solutions; however (iv) is excluded in this case since $\lambda^2$ is real.

As we will see in Sec.~\ref{SecIV}, our numerical results indicate the non-existence of case (ii), that is, we do not find real eigenvalues. In the case of purely imaginary eigenvalues  (case (iii)), we find that $\delta^2\mathcal{E}_{\ell}[ A_R] > 0$ is always positive. In particular, this implies that the ground states have no unstable modes and that such modes locally increase the energy functional $\mathcal{E}_\ell$. For $\ell=0$ this result is in concordance with~\cite{Lieb1977} where it was proven that the $\ell=0$ ground state is a global minimum of $\mathcal{E}_\ell$. For all excited states, we find that case (iv) occurs, implying that they are linearly unstable and correspond to saddle points of $\mathcal{E}_\ell$.

\section{Nonrelativistic $\ell$-boson stars}
\label{SecIII}

As stated in the introduction, numerical solutions of the nonlinear eigenvalue problem~(\ref{SecOrd02lB}) have been given in~\cite{nambo19, jaramillo19} for a few values of $n$ and $\ell$. In this section, we extend the numerical construction to a wider range of $n$ and $\ell$ and discuss some qualitative features of the solutions.

\subsection{Implementation}
\label{IIIA}

\begin{figure*}
	\centering	
    \includegraphics[width=18.cm]{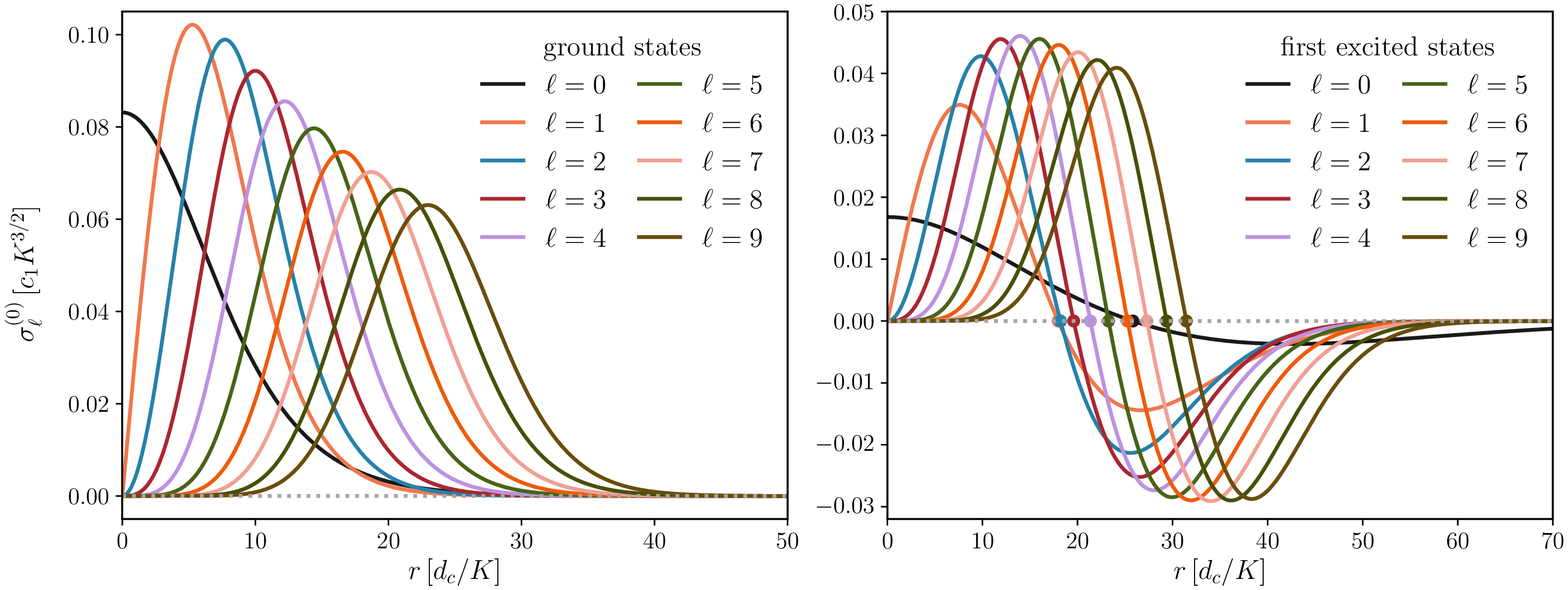}
	\caption{Wave function profiles for different values of $\ell$. Left panel: configurations in the ground state ($n=0$). Right panel: configurations in the first excited state ($n=1$). The thick dots indicate the location of the node in each configuration. Here, $c_1$ refers to the constant $1/\sqrt{(2\ell+1)d_c^3}$ appearing in Eq.~(\ref{AdVar}).}\label{Wfplot}
\end{figure*}

\begin{figure*}
	\centering	
     \includegraphics[width=18.1cm]{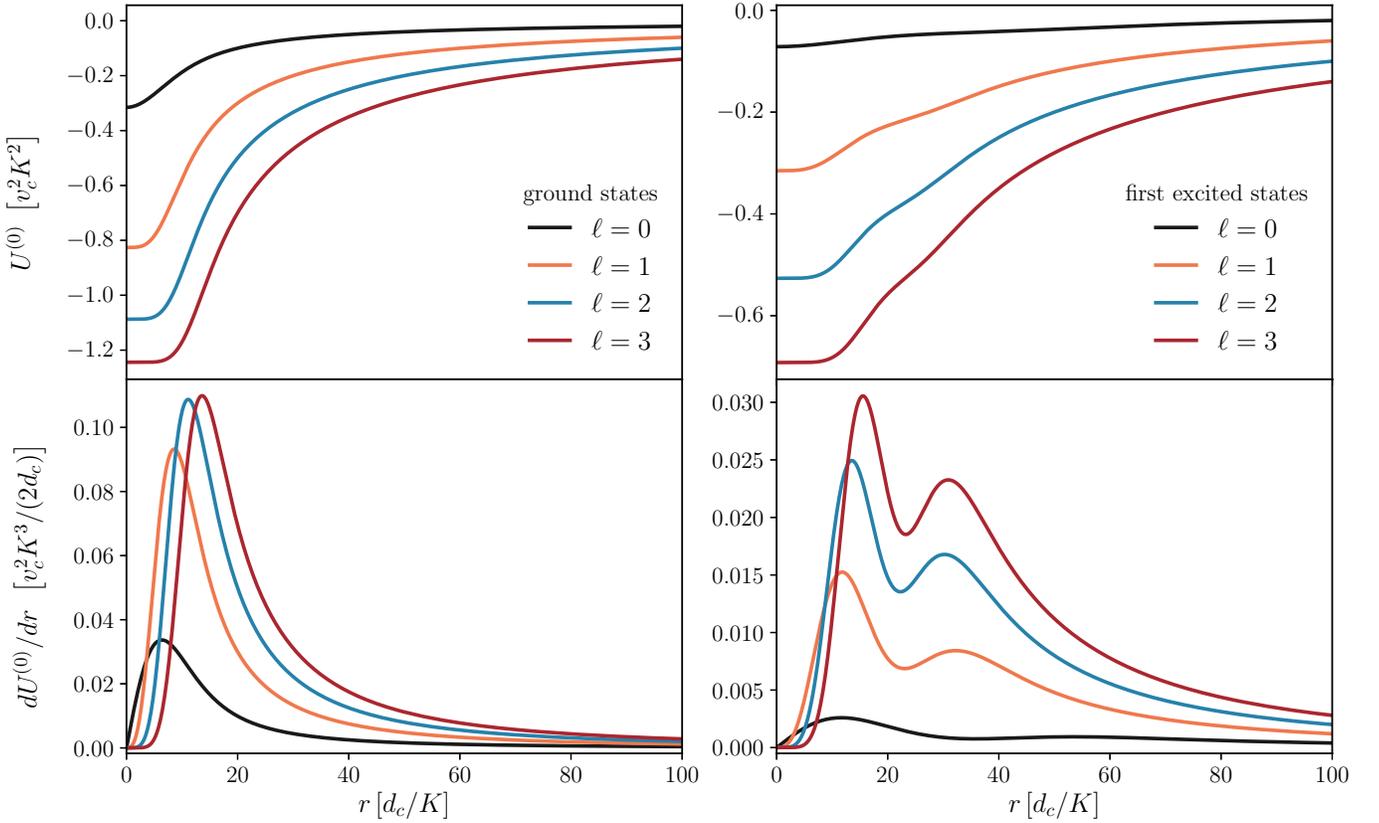}
	\caption{Profiles for the gravitational potential (upper panels) and its first derivative (lower panels) for the configurations shown in Fig.~\ref{Wfplot} corresponding to $\ell = 0,1,2,3$.} \label{Upplot}
\end{figure*}

To perform the numerical integration, it is convenient to replace the gravitational background potential \mbox{$U^{(0)} := \triangle^{-1}_{s}\left(\abs{\sigma_\ell^{(0)}}^2\right)$} with the shifted potential $u^{(0)}(r):= E_\ell - U^{(0)}(r)$, which allows us to rewrite the problem~(\ref{SecOrd02lB}) in the equivalent form
\begin{subequations}\label{Ec2.2.2.1}
\begin{align}
\nabla_s^2 \sigma^{(0)}_{\ell}&= \left[ \frac{\ell(\ell+1)}{r^2}-u^{(0)}\right] \sigma^{(0)}_{\ell},\label{SEc2.2.2.1}\\
\nabla_s^2 u^{(0)}&= - \abs{\sigma^{(0)}_{\ell}}^2.
\end{align}
\end{subequations}

In a next step, we identify the correct boundary conditions at $r=0$ and at $r=\infty$ that guarantee that the solution is normalizable and regular at the origin. Near $r=0$, Eqs.~(\ref{Ec2.2.2.1}) reduce to
\begin{subequations}
\begin{align}
\frac{d}{dr} \left( r^2\frac{d\sigma^{(0)}_{\ell}}{dr}\right)
 -\ell(\ell+1)\sigma^{(0)}_{\ell}&\approx0,\\
\frac{d}{dr}\left( r^2\frac{du^{(0)}}{dr} \right) &\approx 0,
\end{align}
\end{subequations}
whose regular solutions have the form $u^{(0)}(r)\sim \text{const.}$, and $\sigma^{(0)}_{\ell}(r)\sim r^{\ell}$. For this reason, we rescale the wave function as follows:
$\sigma^{(0)}_{\ell}(r) = r^{\ell}\sigma(r)$ with a new radial function $\sigma(r)$ which has a finite nonzero limit as $r\to 0$. By performing a Taylor expansion near $r=0$ one finds the following boundary conditions at the center:\footnote{See~\cite{nambo21} for a rigorous treatment of the local regular solutions near the origin.}
\begin{subequations}\label{BCOr0}
\begin{align}
    \sigma(r=0)&=\sigma_{0},\quad \frac{d\sigma}{dr}(r=0)=0,\\
    u^{(0)}(r=0)& = u_0,\quad \frac{du^{(0)}}{dr}(r=0)=0,
\end{align}
\end{subequations}
with constants $\sigma_0$ and $u_0^{(0)}$. Note that $u_0^{(0)}$ must be positive for a global solution to exist~\cite{nambo21}. Furthermore, by means of the rescaling~(\ref{reescaling}), one can assume without loss of generality that $u_0^{(0)} = 1$. In turn, the value of the constant $\sigma_{0}$ is fine-tuned using a numerical shooting method which aims at the condition $\lim\limits_{r\to\infty}\sigma(r)=0$, which is required for the solution to be normalizable. 

The numerical integration of the system~(\ref{Ec2.2.2.1}) with the boundary conditions~(\ref{BCOr0}) is performed using an adaptive explicit 5(4)-order Runge-Kutta routine\footnote{The integration is performed using the fifth-order accurate steps; the fourth-order steps are only performed in order to estimate the error.}~\cite{2020SciPy-NMeth, DORMAND198019, Lawrence1986SomePR}, where we rewrite the system as a first-order system for the fields $(\sigma,u^{(0)})$. For the fine-tuning, we use a methodology similar to the one described in~\cite{Moroz:1998dh}, based on bisection. Additionally, we find it necessary to match the numerical solution obtained in this way to the asymptotic form of the fields $(\sigma, u^{(0)})$, given by
\begin{subequations}\label{ExtenSol}
\begin{align}
    \sigma(r) &\approx\frac{C_1}{r^{1+\ell-M/(2\kappa)}} e^{-\kappa r},\label{S0Asintotic}\\
    u^{(0)}(r) &\approx E_\ell 
    +\frac{M}{r},\label{u0Asintotic}
\end{align}
\end{subequations}
with $\kappa := \sqrt{|E_\ell|}$ and constants $C_1$, $E_\ell$ and $M$. Here, $E_\ell$ and $M$ represent, respectively, the (unrescaled) energy eigenvalue and total mass of the configuration. The form~(\ref{u0Asintotic}) is obtained by recalling the fact that $\lim\limits_{r\to\infty} U^{(0)}(r)=0$ and the definition $u^{(0)}(r)= E_\ell - U^{(0)}(r)$. Here, the constants $E_\ell$ and $M$ are determined using the methodology described in appendix~\ref{ApendEE}, whereas the constant $C_1$ is computed by fitting the profile of the right-hand side of Eq.~(\ref{S0Asintotic}) to the last $10$ points of the function $\sigma$ obtained from the shooting algorithm. This extension of the solution turns out to be necessary for the numerical analysis of the first-order equations discussed in the next section, which requires the knowledge of the background solution for values of $r$ lying beyond the maximal radius obtained from the shooting algorithm. Our code is publicly available in~\cite{Roque_On_the_radial_2023}. 

The physical energy eigenvalue is obtained as follows. First, $E^{\Lambda}_{\ell}$ is computed using a generalization of the methodology present in~\cite{Moroz:1998dh} (see Appendix~\ref{ApendEE} for more details), according to the formula
\begin{equation}\label{EnergV}
E^{\Lambda}_{\ell}=\frac{(2\ell+1)^2}{M^2} E_\ell,
\end{equation}
where
\begin{subequations}
\begin{align}
E_\ell &= u_{0}-\int_{0}^{\infty} r^{2\ell+1}\abs{\sigma(r)}^2 dr,\\
M &= \int_{0}^{\infty} r^{2(\ell+1)}\abs{\sigma(r)}^2 dr.
\end{align}
\end{subequations}
Next, the dimensional eigenvalue corresponding to $K$ particles in each state $(n, \ell, m)$ is obtained from
\begin{equation}
E^{phys}_{\ell}=2\mu v_c^2 K^2 E_{\ell}^\Lambda,
\label{EnValu}
\end{equation}
and the corresponding total mass is
\begin{equation}
M^{phys} = (2\ell+1) K \mu = N\mu,
\end{equation}
as expected. Alternatively, one can use the asymptotic form described in Eq.~(\ref{u0Asintotic}) to obtain $E_\ell$, assuming that the solution has been properly normalized (as described at the beginning of Sec.~\ref{IIE}) such that $E_\ell = E^{\Lambda}_{\ell}$. This alternative form to compute the energy eigenvalue was used to check the validity of the results obtained from Eq.~(\ref{EnergV}).

\subsection{Results}\label{IIIB}

\renewcommand{\tabcolsep}{11pt}
\renewcommand {\arraystretch}{1.5}
\begin{table*}[t]
	\centering
	\begin{tabular}{c@{\hskip .15 in} r@{\hskip .15 in}| c@{\hskip .2 in} c@{\hskip .2 in} c@{\hskip .2 in} c@{\hskip .2 in}c@{\hskip .2 in}c@{\hskip .2 in}|}
	\cline{3-8}
	\cline{3-8}
		 &&\multicolumn{6}{c|}{
   $E^{phys}_{\ell}\,[K^2 \mu v_c^2]$}\\[-0.1cm]
		 &&\multicolumn{6}{c|}{$\ell$-values:}\\[-0.1cm]
		 && $0$& $1$ & $2$ &$3$ &$4$&$5$ \\[0.1cm]
		\hline
		 \multirow{6}{*}{\begin{sideways}$n$-nodes\end{sideways}}& $0$& $-0.16276921$& $-0.48696445$&$-0.67701587$& $-0.79912216$ & $-0.88467671$ & $-0.9483675$\\[0.1cm]
		 &$1$ &$-0.03079654$& $-0.15836933$&$-0.28505087$& $-0.39233267$ & $ -0.48134634$ &$-0.55563343$\\[0.1cm]
		 &$2$ &$-0.0125261$& $-0.07767856$&$-0.15754598$& $-0.23552556$& $-0.30705008$ & $-0.37133678$\\[0.1cm]
		 &$3$ & $-0.00674732$ & $-0.04597782$ & $-0.09987558$& $-0.15725454$& $-0.21347763$ & $-0.26669641$ \\[0.1cm]
		\hline\hline
	\end{tabular}
	\caption{Energy eigenvalues for the ground and first three excited states computed from Eq.~(\ref{EnergV}) for  $\ell=0,1, \dots, 5$. The eigenvalues obtained for $\ell=0$ and $K=1$ are in agreement with those reported in~\cite{1998MPLA...13.2327B, 2002math.ph...8045H}.\label{tabla1}}
\end{table*}

\renewcommand{\tabcolsep}{11pt}
\renewcommand {\arraystretch}{1.3}
\begin{table*}[t]
	\centering
	\begin{tabular}{c@{\hskip .1 in} r@{\hskip .1 in}| c@{\hskip .1 in} c@{\hskip .1 in} c@{\hskip .1 in} c@{\hskip .1 in}c@{\hskip .1 in}c@{\hskip .1 in}|}
	\cline{3-8}
	\cline{3-8}
		 &&\multicolumn{6}{c|}{$\ell$-values:}\\[-0.1cm]
		 && $0$& $1$ & $2$ &$3$ & $4$ & $5$\\[0.1cm]
		\hline
		 \multirow{4}{*}{\begin{sideways} parameters\end{sideways}}& $\alpha$& $0.0978 \pm 0.0001$& $0.8497 \pm 0.0016$ & $2.1737 \pm 0.0184$& $3.8769\pm0.0637$&$5.8308\pm0.1441$& $7.9516\pm0.2607$ \\[0.1cm]
		 &$\beta$ &$0.7763\pm0.0005$& $1.3222\pm0.0011$&$1.8116\pm0.0057$& $2.2639\pm0.0128$&$2.6907\pm0.0217$& $3.0987\pm0.0320$ \\[0.1cm]
		 &$\gamma$ & $2.0115\pm 0.0007$ & $1.9935\pm0.0011$ & $1.9632\pm0.0040$& $1.9332\pm0.0070$ & $1.9058\pm 0.0098$ &$1.8812\pm0.0123$\\[0.1cm]
		\hline\hline
	\end{tabular}
	\caption{Best fit parameters for the trial function in Eq.~(\ref{EpotL}), taking into account the first thirty energy levels. These parameters are determined using a least square method. Results are shown for angular momenta $\ell=0,1, \dots, 5$.\label{tabla02}}
\end{table*}

\begin{figure*}
	\centering	
    \includegraphics[width=18.cm]{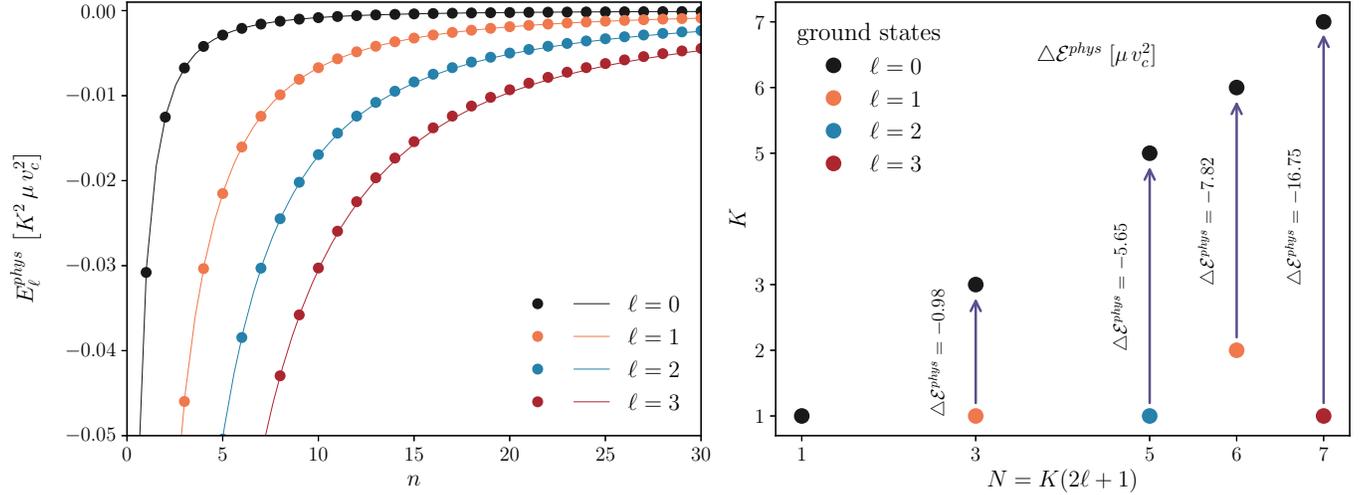}
	\caption{Left panel: numerical results for the first thirty energy levels (shown in the thick dots) corresponding to configurations with angular momenta $\ell=0,1,2,3$. The solid lines show the graph of the trial function in Eq.~(\ref{EpotL}) with the best fit for the parameters $\alpha, \beta, \gamma$ reported in table~\ref{tabla02}. Right panel: energy differences $\triangle \mathcal{E}^{phys}\;[\mu v_c^2]$ computed from the relation Eq.~(\ref{DifferenceEne}) between ground state configurations ($n=0$) with the same numbers of field $N$ but different values of $\ell$, e.g., $\ell=0, K=3$ and $\ell=1, K=1$. The solid arrows indicate the possible loss of energy due to a loss of angular momentum, e.g., $\ell=1 \mapsto \ell=0$.}\label{Enplot}
\end{figure*}

We have solved the system~(\ref{Ec2.2.2.1}) for values of $\ell$ up to $10$ and values of $n$ up to $30$. Typical examples for the radial profile of the wave function for different values of $\ell$ are shown in Fig.~\ref{Wfplot}. The left panel represents configurations in the ground state ($n=0$). Notice that an increase in the value of $\ell$ leads to more flatness near the origin, in accordance with the $r^\ell$-behavior of $\sigma_\ell^{(0)}$, whereas the radius of the maximum increases. This property is in concordance with the corresponding results reported for the relativistic case~\cite{Alcubierre:2018ahf, Alcubierre:2021psa}. In the right panel, we show configurations in the first excited state ($n=1$). Again, higher values of $\ell$ lead to more flatness near the origin and a maximum which lies further away from the origin. The positions of the node and the minimum also move to the right as $\ell$ increases starting from $\ell=1$ (whereas the case $\ell=0$ is special since the field $\sigma_{\ell=0}^{(0)}$ does not vanish at the origin).

Figure~\ref{Upplot} shows the Newtonian potential and its first derivative corresponding to the configurations with $n=0,1$ and $\ell=0,1,2,3$ shown in Fig.~\ref{Wfplot}. As can be seen, the potential profiles are monotonically increasing, which can be understood from the identity $dU^{(0)}/dr = -du^{(0)}/dr$ which is positive according to Eq.~(\ref{du0}) in Appendix~\ref{ApendEE}. We also observe that as $\ell$ increases (and $K$ and $n$ remain fixed), the potential well becomes more profound. This seems to be related to the fact that the number of fields increases with $\ell$ which enhances the gravitational source, although one should be careful with this interpretation since configurations with higher $\ell$ are also more extended as shown in Fig.~\ref{Wfplot}. However, comparing configurations with the same number of fields $N$ and nodes $n$, one finds that the minimum of the potential increases with increasing $\ell$, the deepest well occurring for $\ell=0$. This is the expected behavior one has for a fixed background potential, which is due to the repulsive contribution originating from the centrifugal term in the effective potential (see Eq.~(\ref{SEc2.2.2.1})). Furthermore, as $\ell$ or $n$ increase, the configurations become more extended, which is also visible in the profile of the potential. Note also the existence of an additional local maximum in the derivative of the potential for the first excited state. A parameter exploration seems to indicate that in general, $dU^{(0)}/dr$ has a total of $n+1$ maxima, corresponding to the $n+1$ maxima of the density $|\sigma_\ell^{(0)}|^2$.

Table~\ref{tabla1} shows the results for the energy levels corresponding to the ground and first three excited states for $\ell=0, 1, 2, 3, 4, 5$, computed using Eq.~(\ref{EnergV}). The left panel of Fig.~\ref{Enplot} shows the energy levels as a function of $n$ for $\ell=0,1,2,3$. As commented in the introduction, for each fixed value of $\ell$, the eigenvalues are negative and increase monotonically in $n$. Similar to what has been found in the case $\ell=0$~\cite{1998MPLA...13.2327B}, this increase can be fitted to an inverse power law of the form
\begin{align}
     E^{phys}_{\ell} =-\frac{\alpha}{(n+\beta)^{\gamma}}\times \,[K^2\mu v_c^2], \label{EpotL}
\end{align}
with suitable parameter values for $\alpha$, $\beta$ a $\gamma$. Table~\ref{tabla02} shows the best fit for these parameters using a least square method, for the cases $\ell=0, 1, 2, 3, 4, 5$. The resulting curve is also shown in the left panel of Fig.~\ref{Enplot}. Remarkably, the value for the exponent $\gamma$ seems to lie quite close to $2$. In fact, by performing the fit using only the excited states $n = 10, 11, \ldots, 30$ one finds values of $\gamma$ which are consistent with $2$ up to a relative error of $0.4\%$, at least for the values of $\ell = 0, 1, \ldots, 5$ reported in the table. Therefore, for large values of $n$ and fixed $\ell$, $E^{phys}_{\ell}$ seems to behave as $-\alpha/n^2$, similar to the Balmer spectrum arising from the Coulomb potential $\mathbf{U} = -G M/r$. However, note that in the Coulomb case, the parameter $\beta$ would be equal to $\ell+1$, which is clearly not the case here. This is probably related to the fact that although in our configurations $\mathbf{U}$ behaves as the Coulomb potential in the asymptotic region $r\to\infty$, it is regular at the center.

Next, we compare the total energy $\mathcal{E}^{phys}_{\ell}$ of different configurations having the \textit{same number} of fields. For this, recall that the nonrelativistic $\ell$-boson stars are composed of $N=K(2\ell+1)$ self-gravitating bosons. Unlike the standard nonrelativistic boson stars which have $N=1$, considering configurations with $N\geq 3$ offers the interesting possibility of constructing configurations with the same $N$ but different values for $K$ and $\ell$, like for instance in the pairs $(K, \ell) = (3, 0)$ and $(K, \ell) = (1, 1)$. A relevant question is which of these configurations has the least total energy $\mathcal{E}^{phys}_{\ell}$ since this is expected to be the most stable state (at least within the spherically symmetric configurations). Using Eqs.~(\ref{Eq:FunctionalValueLBS}, \ref{EnValu}), the difference between two configurations $(K_1,\ell_1)$ and $(K_2,\ell_2)$ with $N = K_1(2\ell_1+1) = K_2(2\ell_2+1)$ fields is
\begin{align}
   \triangle \mathcal{E}^{phys}_{1\to 2}=&\mathcal{E}^{phys}_{2}-\mathcal{E}^{phys}_{1},\nonumber\\
    =&\frac{K_1^{2}N}{3}\left[ \frac{\left(2\ell_1+1\right)^2}{\left(2\ell_2+1\right)^2} E_2 -E_{1}\right],\label{DifferenceEne}
\end{align}
where the energies $E_i$ refer to the physical energy levels $E^{phys}_{\ell}$ for $K=1$ in both configurations (i.e., the values reported in Table~\ref{tabla1} in units of $\mu v_c^2$). For excited states, one may also use Eq.~(\ref{EpotL}) with the associated parameters from Table~\ref{tabla02} to provide approximate values for $E_1$ and $E_2$ and compute the energy difference.

The right panel of Fig.~\ref{Enplot} shows the energy difference between ground state $\ell$-boson stars with $\ell=\ell_1 > 0$ and ground state $\ell$-boson stars with $\ell = \ell_2 = 0$ and the same number of particles for the cases $(K_1, \ell_1) = (1, 1)$, $(2, 1)$, $(1, 2)$, $(1, 3)$. As can be appreciated from this plot, this energy difference is always negative, $\triangle \mathcal{E}^{phys}_{1\to 2} < 0$ and becomes larger as the value of $\ell_1$ increases. A more general exploration based in the fitting formula~(\ref{EpotL}) reveals the following properties: when $n_1\geq n_2$ and $\ell_1\geq\ell_2$, one has $\triangle \mathcal{E}^{phys}_{1\to 2} < 0$, as expected. This is consistent with the fact that for any given number of wave functions, the configuration corresponding to $K=N$, $n=\ell=0$ represents the global minimum of the conserved energy functional Eq.~(\ref{ecFuncellBS}), as pointed out in the introduction. Regarding the energy difference between configurations with $n_1\geq n_2$ and $\ell_1 < \ell_2$, it turns out it can also be negative, provided $n_1$ is sufficiently large. An example is provided by the cases of $N=9$ fields and $n_1\geq 2, \ell_1=1$, $n_2=0$ and $\ell_2=4$ (whereas in this example $\triangle \mathcal{E}^{phys} > 0$ if $n_2 = 2$ but $n_1 = 1$ or $n_1=2$).

\section{Linear stability of nonrelativistic $\ell-$boson stars}
\label{SecIV}

In the previous section we constructed nonrelativistic $\ell-$boson stars and discussed their main properties. In this section we present our numerical implementation and main results corresponding to their linear stability.

\subsection{Linear system, boundaries conditions and discretization}\label{IVA}

Similar to the background equations, for the numerical implementation of the linearized system~(\ref{SecOrd12lB}), it is convenient to rewrite it in a more appropriate form. For this, we rewrite $A(r):=a(r)/r$, $B(r):= b(r)/r$ with rescaled functions $a$ and $b$, and use the identity $\laplacian_s = \frac{1}{r}\frac{d^2}{dr^2} r$ to rewrite this system as
\begin{subequations}\label{Ec2.2.2.2}
\begin{align}
b'' - U_{\text{eff}}\,b&=-i\lambda a, \\
a'' - U_{\text{eff}}\, a
 -2\sigma^{(0)}_{\ell}\left( \frac{d^2}{dr^2} \right)^{-1}\left[\sigma_\ell^{(0)} a \right] &=-i\lambda b,\label{Eq48b}
\end{align}
\end{subequations}
where for convenience we have introduced the effective potential $U_{\text{eff}}(r):= -u^{(0)}(r) + \ell(\ell+1)/r^2$ and the operator $\left( \frac{d^2}{dr^2} \right)^{-1} = r\triangle_s^{-1} r^{-1}$ denoting the inverse of the second derivative with homogeneous Dirichlet conditions at $r=0$ and $r=\infty$.

To solve the system~(\ref{Ec2.2.2.2}), four  boundary conditions are needed. Similar to the analysis applied to the background configurations, one can study (heuristically) the dominant terms of the perturbed system near the origin and infinity. Using the fact that $\sigma_\ell^{(0)}(r)\sim r^{\ell}$, one finds
\begin{align}
r^2 Z''-\ell(\ell+1) Z\approx0, \nonumber
\end{align}
near $r=0$, with the column vector $Z:=(a,b)^{T}$. The solution which is regular at the center behaves as $Z(r)\sim r^{\ell+1}$, which leads to the following boundary conditions for all $\ell\geq 0$ at the origin:
\begin{subequations}\label{PertBC}
\begin{align}
a(r=0)=0,\qquad b(r=0)=0.
\end{align}
In the asymptotic region one finds, taking into account the fact that $u^{(0)}(r)\to E_\ell$ and that $\sigma_\ell^{(0)}$ decays exponentially,
\begin{align}
Z'' + E_\ell Z\approx0, \nonumber
\end{align}
and the solution that is bounded at infinity is the one that decays exponentially. Hence, as $r\to \infty$, we require that
\begin{align}
\lim\limits_{r\to\infty} a(r)=0,\qquad \lim\limits_{r\to\infty} b(r)=0.
\end{align}
\end{subequations}

In order to numerically solve the linearized system~(\ref{Ec2.2.2.2}) we used the background solutions $(\sigma_\ell^{(0)}, u^{(0)})$ found in the previous section, and we represent these, as well as the perturbed fields $(a,b)$, in terms of Chebyshev polynomials. The derivative operators are discretized using a standard spectral method (see, e.g.,~\cite{trefethen2000spectral, boyd2013chebyshev}), which leads to a matrix eigenvalue problem. The next paragraph briefly describes the details of this implementation. 

First, we map the domain $\mathsf{D_{C}}:= [-1, 1]$, on which the Chebychev polynomials are defined, onto the physical domain $\mathsf{D}:=[0, r_{\star}]$ which is truncated at a large radius $r_\star$ (in Appendix~\ref{apend2} we also consider the case $r_\star\to \infty$ in which the whole physical domain $[0,\infty)$ is covered). Specifically, we define this map $\mathsf{D_C}\to \mathsf{D}$ through the transformation, $r=r_{\star}(x+1)/2$ with $x\in \mathsf{D_C}$~\cite{2002math.ph...8045H}. Second, on $\mathsf{D_C}$ we introduce the set of Chebyshev points $x_j=\cos(j\pi/\mathsf{N})$, $j = 0, 1,\dots, \mathsf{N}$, and we discretize $d/dx$ using the Chebyshev differentiation matrix $\mathbb{D}_{\mathsf{N}}$. Since $dr/dx = r_\star/2$, the corresponding discretization of the second derivative operator $d^2/dr^2$ yields $(2/r_{\star})^2\mathbb{D}^2_{\mathsf{N}}$. The explicit form of the $(\mathsf{N}+1)\times (\mathsf{N}+1)$ matrix $\mathbb{D}_{\mathsf{N}}$ can be found in Chapter 6 of~\cite{trefethen2000spectral}. Third, in order to impose the homogeneous boundary conditions~(\ref{PertBC}) we use the procedure described in~\cite{trefethen2000spectral} which amounts in striking the first and last rows and columns in the second derivative differentiation operator $\mathbb{D}_{\mathsf{N}}^{2}$, giving rise to an $(\mathsf{N}-1)\times (\mathsf{N}-1)$-matrix. This reduced matrix is then inverted in order to discretize the operator $(d^2/dr^2)^{-1}$ appearing in Eq.~(\ref{Eq48b}). 

Using everything previously mentioned, the problem~(\ref{Ec2.2.2.2}) is reduced to the finite-dimensional eigenvalue problem
\begin{widetext}
\begin{gather}\label{Syst}
 \begin{pmatrix} \mathsf{0} & \mathbb{\tilde{D}}_{\mathsf{N}}^{2}-U_{\text{eff}} \\ \mathbb{\tilde{D}}_{\mathsf{N}}^{2}-\mathsf{U}_{\text{eff}}-2\Sigma_{0}\big(\mathbb{\tilde{D}}_{\mathsf{N}}^{2}\big)^{-1} \Sigma_0 & \mathsf{0}\end{pmatrix}\begin{pmatrix} \mathsf{a} \\ \mathsf{b}\end{pmatrix}
 =
 -i\lambda
   \begin{pmatrix} \mathsf{a} \\ \mathsf{b}\end{pmatrix},
\end{gather}
\end{widetext}
where here $\mathsf{0}$ represents the $(\mathsf{N}-1)\times (\mathsf{N}-1)$ zero matrix,\newline\\

\begin{align}
U_{\text{eff}}&:=\textbf{diag}\bigg(U_{\text{eff}}(x_1), U_{\text{eff}}(x_2),\dots, U_{\text{eff}}(x_{\mathsf{N}-1})\bigg),\nonumber\\
\Sigma_0&:=\textbf{diag}\bigg(\sigma^{(0)}_{\ell}(x_1),\sigma^{(0)}_{\ell}(x_2),\dots,\sigma^{(0)}_{\ell}(x_{\mathsf{N}-1})\bigg),\nonumber
\end{align}
are the discrete representation of the background quantities $U_{\text{eff}}$ and $\sigma_\ell^{(0)}$ and the vector
\begin{align}
\begin{pmatrix} \mathsf{a} \\ \mathsf{b}\end{pmatrix}
:=\bigg( a(x_1), \;\dots\; ,a(x_{\mathsf{N}-1}), b(x_1),\dots, b(x_{\mathsf{N}-1})\bigg)^{T},\nonumber
\end{align}
represents the eigenfields $r(A,B)^T$. We solve the discrete eigenvalue problem~(\ref{Syst}) using the SciPy library~\cite{2020SciPy-NMeth}. After some experimentation, we have found that using the number $\mathsf{N}:=3r_{\star}/4$ of Chebyshev points\footnote{Recall that these points are not uniformly distributed; the density of points is largest near the boundaries of $\mathsf{D_C}$.} with outer boundary located at $r_{\star}:=200(n+1)$ for the $n$'th excited state of the background solution, gave accurate results (see Appendix~\ref{apend2} for details). Since $r_\star$ is much larger than the typical maximal radius obtained from the shooting algorithm, we extend the background solution on $[0, r_{*}]$ using the asymptotic expressions~(\ref{ExtenSol}), as described in Sec.~\ref{IIIA}. A validation of our results which is based on a convergence study of the numerical solution as $\mathsf{N}$ and $r_\star$ vary, as well as on an independent residual evaluation check using an explicit Runge-Kutta method are provided in Appendix~\ref{apend2}.

\subsection{Mode stability of the linear system}\label{IVB}

\renewcommand{\tabcolsep}{10pt}
\renewcommand {\arraystretch}{1.5}
\begin{table*}[t]
	\centering
	\begin{tabular}{c@{\hskip .1 in} r@{\hskip .08 in}| c@{\hskip .1 in} c@{\hskip .1 in} c@{\hskip .1 in} c@{\hskip .05 in}|}
	\cline{3-6}
	\cline{3-6}
		 &&\multicolumn{4}{c|}{
   $\lambda\,[K^2/t_c]$}\\[-0.1cm]
		 &&\multicolumn{4}{c|}{$\ell$-values:}\\[-0.1cm]
		 && $0$& $1$ & $2$ &$3$ \\[0.1cm]
		\hline
		 \multirow{10}{*}{\begin{sideways}$n$-nodes\end{sideways}}& \multirow{3}{*}{$0$}& $0\pm 0.03412558i$& $0\pm 0.06385090i$&$0\pm0.06408695i$& $0\pm0.05903587i$\\[-0.2cm]
		 & &$0\pm 0.06030198i$ &$0\pm 0.14328038i$ & $0\pm 0.16575905i$& $0\pm0.16827469i$\\[-0.2cm]
		 & &$0\pm 0.06882477i$ & $0\pm 0.17565513i$ & $0\pm 0.21233555i$& $0\pm0.22189706i$\\[0.1cm]
		 &\multirow{3}{*}{$1$} &$0\pm 0.00300045i$& $0\pm0.01185117i $& $0\pm 0.01736224i$& $0\pm0.02017560i$ \\[-0.2cm]
		 & &$0\pm 0.00812230i$ & $0\pm 0.04155775i$ & $0\pm0.06711048i$& $0\pm0.08323611i$\\[-0.2cm]
		 & &$0\pm 0.01073335i$ & $0\pm 0.05025166i$ & $0\pm 0.08170887i$& $0\pm0.10244472i$\\[0.1cm]
		 &\multirow{3}{*}{$2$} &$0 \pm 0.00078452i $& $0\pm0.00406666i$ &$0\pm0.00709575i$& $0\pm0.00931852i$\\[-0.2cm]
		 & &$0 \pm 0.00297490i $& $0\pm0.01667870i $&$0\pm0.03118429i$& $0\pm0.05061671i$\\[-0.2cm]
		 & &$0 \pm 0.00366760i $& $0\pm0.02154496i$&$0\pm 0.04306049i$& $0\pm0.05994153i$\\[0.1cm]
		 &\multirow{3}{*}{$3$} &$0\pm 0.00031089i$& $0\pm0.00185072i$&$0\pm0.00357142i$& $0\pm0.00506232i$\\[-0.2cm]
		 & &$0\pm 0.00134987i$& $0\pm0.00870332i$&$0\pm0.01961872i$& $0\pm0.02916710i$\\[-0.2cm]
		 & &$0\pm 0.00182670i$& $0\pm0.01175539i$&$0\pm 0.02394986i$& $0\pm0.03932200i$\\[0.1cm]
		\hline\hline
	\end{tabular}
	\caption{First three purely imaginary eigenvalue pairs for the background configurations with $n=0,1,2,3$ and $\ell=0,1,2,3$.
	\label{tabla_imaginary}}
\end{table*}

\begin{figure*}
	\centering	
	\includegraphics[width=18.cm]{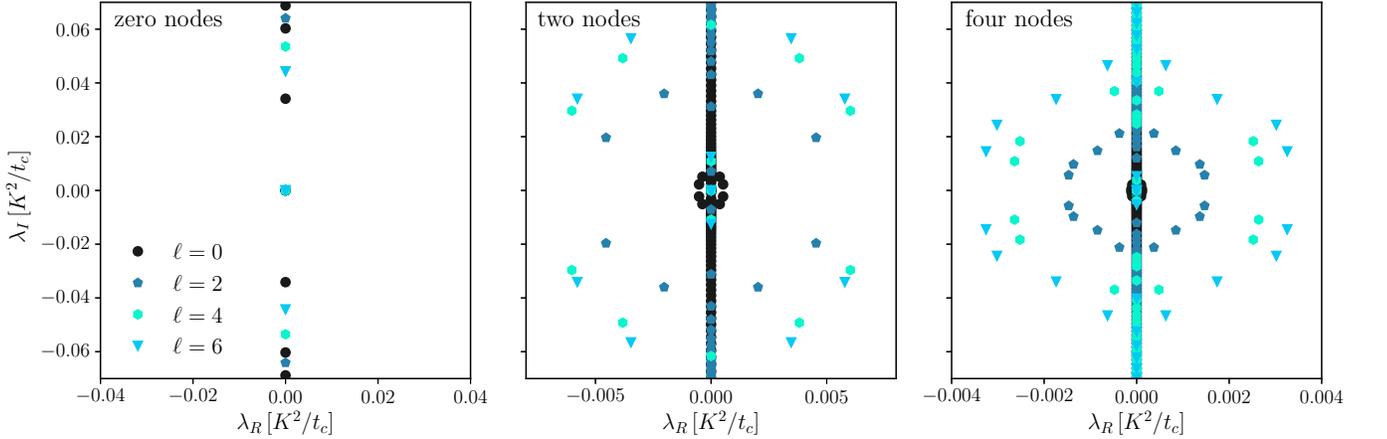}
	\caption{Eigenvalue spectra for the ground state configurations $n=0$ (left panel) and the excited states with $n=2$ (center panel) and $n=4$ (right panel) for $\ell = 0,2,4,6$. The points $\lambda=0$ correspond to the zero modes describing a phase change of the background solution (see subsection~\ref{IIE} for more details). Notice that the eigenvalues come in quadruples $\{\lambda, -\lambda, \lambda^{*},-\lambda^{*}\}$, as discussed in subsection~\ref{IIE}. The only background configurations for which the spectrum is purely imaginary (and therefore leads to linear stability) are the ground states  with zero nodes; the remaining states with $n\geq 1$ nodes have precisely $n-$ quadruple eigenvalues with non-zero real parts which imply that they are linearly unstable.\label{Lplot}}
\end{figure*}

\renewcommand{\tabcolsep}{10pt}
\renewcommand {\arraystretch}{1.5}
\begin{table*}[t]
	\centering
	\begin{tabular}{c@{\hskip .1 in} c@{\hskip .1 in}| l@{\hskip .15 in}| l@{\hskip .2 in}|l@{\hskip .2 in}||}
	\cline{3-5}
	\cline{3-5}
		 $\ell$ & $n$ & & &  \\[-0.25cm]
		 values & nodes & \multicolumn{1}{|c|}{$\lambda\,[K^2/t_c]$} & $\delta^2\mathcal{E}_{\ell}[A_{R}]$ & $\delta^2\mathcal{E}_{\ell}[A_{I}]$ \\[0.2cm]
		 \hline
		 \multirow{3}{*}{$0$} & $0$ & $0\pm 0.03412558i$ & $0.00637265$ & $0$ \\[0.1cm]
		 & \multirow{2}{*}{$1$} & $0\pm 0.00300045i$ & $0.00039739$  & $0$ \\[-0.2cm]
		 && $\pm 0.00148347 +0.00979587i$ & $0.00011228$ & $-0.00011228$\\[0.1cm]		 \hline
		 \multirow{3}{*}{$1$} & $0$ & $0\pm 0.06385090i$ & $0.00787292$ & $0$ \\[0.1cm]
		 & \multirow{2}{*}{$1$} &$0\pm0.01185117i$ & $0.00098164$ & $0$\\[-0.2cm]
		 && $\pm0.00619398 + 0.03340186i$ & $0.00028039$ & $-0.00028043$ \\[0.1cm]
		\hline\hline
	\end{tabular}
	\caption{Numerical values of the second variation $\delta^2\mathcal{E}_{\ell}[A_{i}]$, $i=R,I$, of the conserved energy functional computed using Eqs.~(\ref{EqB3}). The results shown correspond to the first few eigenvalues and eigenfunctions associated with the background solutions with $\ell=0, 1$ and $n=0, 1$.\label{tabla0}
}
\end{table*}

\renewcommand{\tabcolsep}{10pt}
\renewcommand {\arraystretch}{1.5}
\begin{table*}[t]
	\centering
	\begin{tabular}{c@{\hskip .1 in} r@{\hskip .08 in}| c@{\hskip .1 in} c@{\hskip .1 in} c@{\hskip .1 in} c@{\hskip .05 in}|}
	\cline{3-6}
	\cline{3-6}
		 &&\multicolumn{4}{c|}{$\lambda\,[K^2/t_c]$}\\[-0.1cm]
		 &&\multicolumn{4}{c|}{$\ell$-values:}\\[-0.1cm]
		 && $0$& $1$ & $2$ &$3$ \\[0.1cm]
		\hline
		 \multirow{7}{*}{\begin{sideways}$n$-nodes\end{sideways}}& $0$& $-$& $-$&$-$& $-$\\[0.1cm]
		 &$1$ &$0.00148347+0.00979587i$& $0.00619398+0.03340186i$& $0.00888859+0.04742445i$&$0.00923822+0.05471082i$\\[0.1cm]
		 &\multirow{2}{*}{$2$} &$0.00037420+0.00507753i$&$0.00135751+0.02218741i$&$0.00202943+0.03598465i$&$0.00282615+0.04277090i$\\[-0.2cm]
		 & &$0.00051995+0.00225911i$& $0.00271729+0.01142125i$&$0.00454164+0.01963532i$& $0.00559009+0.02557525i$\\[0.1cm]
		 &\multirow{3}{*}{$3$} &$0.00014893+0.00309572i$& $0.00057487+0.01547544i$&$0.00094498+0.02698601i$& $0.00107542+0.03559566i$\\[-0.2cm]
		 & &$0.00017445+0.00168940i$& $0.00086410+0.00955161i$&$0.00177133+0.01701821i$& $0.00273695+0.02361055i$\\[-0.2cm]
		 & &$0.00022483+0.00088225i$& $0.00132314+0.00519695i$&$0.00247179+0.00993824i$& $0.00335300+0.01399112i$\\[0.1cm]
		\hline\hline
	\end{tabular}
	\caption{Eigenvalue quadruples $\{\lambda, -\lambda, \lambda^{*},-\lambda^{*}\}$ which have real parts different from zero for the background configurations with $n=0,1,2,3$ and $\ell = 0,1,2,3$. Only the member with positive real and imaginary parts is reported. The corresponding spectrum in the complex plane is illustrated in Fig.~\ref{Lplot} for the configurations with $n=0,2,4$ and $\ell=0,2,4,6$.\label{tabla2}}
\end{table*}

\begin{figure*}
	\centering	
	\includegraphics[width=17.5cm]{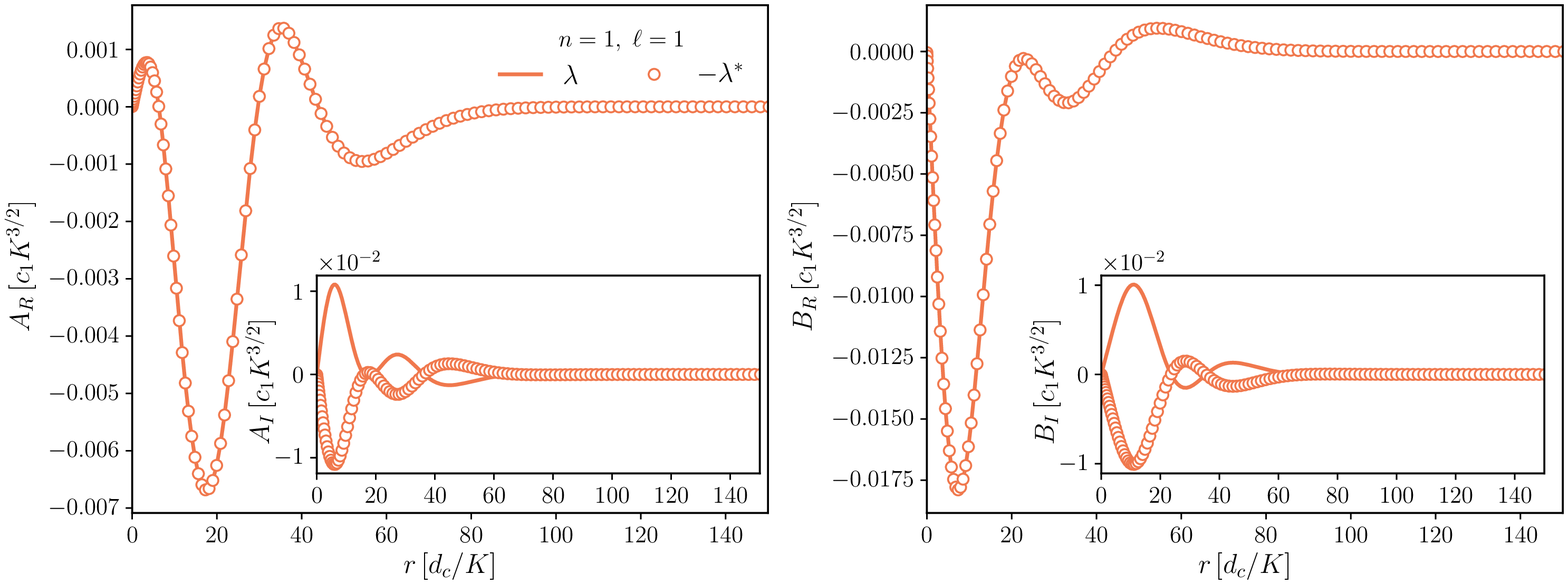}
	\caption{Eigenfuncion profiles for the fields $A$ (left panel) and $B$ (right panel) for the first excited background configuration with $\ell=1$. Shown is the eigenvalue-eigenfunction pair $\{ (\lambda,A,B), (-\lambda^*,A^*,B^*) \}$ (see Eq.~(\ref{Eq:lambda4})) corresponding to $\lambda = 0.00619398 + 0.03340186i$ reported in table~\ref{tabla2}. The exterior plots show the real parts of $A$ and $B$ while the plots in the inset show their imaginary parts. Notice that $A_R$ and $B_R$ agree with each other, while $A_I$ and $B_I$ differ from each other by a sign, as expected from the above-mentioned symmetry $(\lambda, A, B)\mapsto (-\lambda^*, A^*, B^*)$. Here, $c_1$ refers to the constant $1/\sqrt{(2\ell+1)d_c^3}$.}\label{Efplot}
\end{figure*}

After having described our numerical implementation, we turn our attention to the main results of this section, which reveals the behavior of the mode solutions of the linear perturbations of the nonrelativistic $\ell$-boson stars. To this purpose, we recall the general form of these modes in Eq.~(\ref{Ecpert}).  Separating the fields $(A, B)$ and the eigenvalue in their real and imaginary parts, this equation can be rewritten as
\begin{align}
\label{EqMod}
\sigma_\ell(t,r) &= 2e^{\lambda_R t}\cos{(\lambda_{I}t)}\left[A_R(r) + i B_I(r)\right]\nonumber\\
&-2e^{\lambda_R t}\sin{(\lambda_{I}t)}\left[ A_I(r) - i B_R(r)\right],
\end{align}
where we recall that the sub-indices $R$ and $I$ refer to the real and imaginary parts, respectively. As is evident from this equation, a positive value of $\lambda_R$ implies that the mode is exponentially growing, whereas $\lambda_I$ describes its oscillation frequency.

Similar to subsection~\ref{IIE} we divide the stability study into two parts. First, we study the modes corresponding to the ground configurations and next the ones associated with the excited states.

\subsubsection{Ground state configurations}

As we proved in subsection~\ref{IIE}, for the ground state configurations, $\lambda^2$ must be real, which implies that $\lambda$ itself is either real or purely imaginary. As a first result of our numerical study, we have found that only purely imaginary values of $\lambda$ occur for the ground states.\footnote{More precisely, we have found that the eigenvalues computed from the $2(\mathsf{N}-1)\times 2(\mathsf{N}-1)$ matrix in Eq.~(\ref{Syst}) have real parts which are smaller than $10^{-6}$ in magnitude.} The first eigenvalues (ordered according to their magnitude) are shown in the first three rows of Table~\ref{tabla_imaginary} for $\ell = 0, 1, 2, 3$. The left panel of Fig.~\ref{Lplot} shows the spectrum in the complex plane corresponding to the ground state configurations with $\ell = 0, 2, 4, 6$. These results are compatible with the relation~(\ref{Eq:lambda4}), which implies that the purely imaginary eigenvalues come in pairs $\{ \lambda,-\lambda \}$.

Since the ground state configurations only have purely imaginary eigenvalues, we conclude from Eq.~(\ref{EqMod}) that the configurations with $n=0$ and the values of $\ell$ analyzed only possess purely oscillatory modes (whose frequency is equal to $ \lambda_{I}$). \emph{This leads us to conjecture that all ground state background configurations are stable with respect to time-dependent linear perturbations of the form~(\ref{eq:ansatzPert}, \ref{Ecpert}).}
Furthermore, since $-\lambda^2 > 0$, we conclude that all these modes give rise to a positive second variation of the conserved energy functional: $\delta^2\mathcal{E}_{\ell}[A_R] > 0$, see subsection~\ref{IIE}. (Table~\ref{tabla0} shows an example in which the second variation is computed numerically and its sign is found to agree with the one of $-\lambda^2$, as expected.) This implies that the ground states represent local minima of $\mathcal{E}_{\ell}$ with respect to such perturbations. However, recall from the results in the previous section that ground state configurations with $\ell > 0$ have values of $\mathcal{E}_{\ell}$ lying above the corresponding value of $\mathcal{E}_{\ell=0}$ with $N=(2\ell+1)K$ fields, implying that these states cannot represent a \emph{global} minimum of the energy functional. \emph{Therefore, we conclude that ground state configurations with $\ell>0$ are either local minima or saddle points of the conserved energy functional $\mathcal{E}_\ell$ under arbitrary perturbations (with or without symmetries).}

\subsubsection{Excited states}

Next, we turn our attention to the mode stability of the background configurations for which $n > 0$. Recall the four possibilities (i-iv) discussed in subsection~\ref{IIE}. First, let us explain the significance of these four cases for the behavior of the corresponding mode described by Eq.~(\ref{EqMod}). Clearly, case (i) corresponds to a time-independent perturbation since $\lambda=0$. As discussed above, it describes an infinitesimal phase change of the background solution. Next, case (ii) corresponds to a pair of modes, one growing exponentially in time and the other one decaying exponentially. Thus, its occurrence would imply that the underlying background solution is linearly unstable. Next, case (iii) describes a pair of purely oscillatory modes with frequency $\lambda_{I}\neq 0$.
Finally, case (iv) gives rise to a quadruple of oscillating modes, two of which have an exponentially growing amplitude, whereas the other two have a decaying amplitude. The occurrence of this case implies that the background solution is linearly unstable and that it represents a saddle critical point of the conserved energy functional, provided the non-degeneracy condition $\delta^2\mathcal{E}_{\ell}[A_R]\neq 0$ holds.

Figure~\ref{Efplot} shows an illustrative example for the eigenfunction profiles of the fields $(A, B)$ for the two eigenvalues $\lambda=0.00619398+0.03340186i$ and $-\lambda^{*}$ reported in table~\ref{tabla2} corresponding to the first excited state $n=1$ with $\ell=1$. Note the linear behavior near the center (which is compatible with the asymptotic behavior $a(r), b(r)\sim r^{\ell+1}$ at the origin) and the decay of the amplitude for large radii, which is compatible with the exponential decay of the fields at infinity. Note also the relation between the eigenfields corresponding to $\lambda$ and $-\lambda^*$, which is in agreement with Eq.~(\ref{Eq:lambda4}).

Our eigenvalue analysis reveals the following. First, as for the ground state configurations, we found purely imaginary eigenvalues, corresponding to case (iii). The first few of them are exhibited in Table~\ref{tabla_imaginary} for $n=1,2,3$ and $\ell=0,1,2,3$. However, in contrast to the ground states, we have also found quadruples of eigenvalues with $\lambda_{R}\neq 0$ and $\lambda_{I}\neq 0$, corresponding to case (iv), implying that the underlying configurations are linearly unstable. Interestingly, our results indicate that in each case, there are precisely $n$ of such quadruples, with $n$ the number of nodes of the background solution (see the center and right panels of Fig.~\ref{Lplot} and Table~\ref{tabla2} for some examples). Furthermore, our results seem to indicate that for each of these quadruples, the non-degeneracy condition $\delta^2\mathcal{E}_{\ell}[A_R]\neq0$ is satisfied (see table~\ref{tabla0} for specific examples with $n=1$ and $\ell=0,1$). The presence of these $n$ quadruples implies that the excited states represent saddle points of the energy functional $\mathcal{E}_{\ell}$. Finally, our results indicate that case (ii) never occurs since we have not found any purely real eigenvalues aside from the zero mode.
\emph{These results lead us to the conjecture that all configurations $(n,\ell)$ with $n\geq 1$ and $\ell\geq 0$ are linearly unstable and possess precisely $2n$ exponentially in time growing modes of the form~(\ref{eq:ansatzPert}, \ref{Ecpert}). Furthermore, these configurations  correspond to saddle critical points of the energy functional.}

Before concluding this section, we would like to make a few remarks regarding the behavior of the eigenvalues $\lambda$ that can be inferred from Tables~\ref{tabla_imaginary} and~\ref{tabla2} and further data corresponding to higher values of $n$ and $\ell$ which are not shown in these tables. First, let us analyze the period of the first purely oscillatory modes (i.e., the slowest oscillating one for each $\ell$ and $n$) in Table~\ref{tabla_imaginary}. We observe that this period increases with $n$ for fixed $\ell$, the shortest period belonging to the ground state. Second, let us compare the real parts of $\lambda$ in a given column in Table~\ref{tabla2} (i.e., fixing $\ell$) for different values of $n$. Interestingly, the shortest living unstable mode for each $n$ (i.e., the one with the largest real part) has a lifetime that \emph{increases} with $n$. In this sense, higher excited states are ``less unstable" than lower excited ones. Third, let us perform the same comparison for a fixed row  in Table~\ref{tabla2} (i.e., a fixed number $n$ of nodes) and configurations with different values of $\ell$. This comparison can be performed either for fixed $K$ or for fixed number $N = K(2\ell+1)$ of fields. In the first case, the shortest living mode has a lifetime that decreases with increasing $\ell$. In the second comparison, the opposite occurs, i.e., configurations with higher $\ell$'s are less unstable.  

\section{Conclusions}
\label{SecV}

We started this article by considering a non-relativistic system of $N$ identical particles with zero spin, interacting only through the common gravitational potential they generate. By writing the $N$-particle wave function as a symmetrized product of one-particle states whose angular dependency has the particular form~(\ref{ansatzAng}), the system was reduced to the effective one-dimensional SP system~(\ref{ellSis}). Stationary solutions of this system describe nonrelativistic $\ell$-boson stars which are compact objects generalizing the standard $\ell=0$ configurations by extending the internal symmetry group from $U(1)$ to $U(N)$. These objects are characterized by the numbers $(N,\ell,n)$ with $\ell$ representing the angular momentum number of the fields, $n$ the node number of the radial wave function and $N$ being equal to an integer multiple of $2\ell+1$. For fixed values of $N$ and $\ell$ their energy levels grow monotonically with $n$, the ground state $n=0$ having minimum energy. 

However, for the stability properties of the nonrelativistic $\ell$-boson star configurations, it is the total energy (which includes the gravitational binding energy in addition to the energy of the wave functions) that turns out to be more relevant than their energy levels. As we have shown in Appendix~\ref{App:LagrangeFormulation}, a conserved energy functional can be naturally derived from the Lagrangian formulation of the SP system. The rescaling freedom (see Eq.~(\ref{reescaling})), through Noether's theorem, gives rise to the connection formula~(\ref{Eq:FunctionalValueLBS}) between the stationary configurations' total energy and their energy levels. 

Although these results constitute a straightforward generalization of known results for the $\ell=0$ case, the inclusion of the angular momentum $\ell$ leads to interesting new effects. For instance, configurations with $\ell > 0$ have zero density at their center and thus -- like their relativistic counterparts~\cite{Alcubierre:2018ahf, Alcubierre:2021psa} -- their morphology is shell-like, where the shell's radius increases with $\ell$ (see Fig.~\ref{Wfplot}). Regarding the stability property, recall that for each fixed value of $N$ the configuration $(N,0,0)$ represents the global minimum of the conserved energy functional and thus is expected to be stable with respect to small enough perturbations. Therefore, the question arises whether or not configurations $(N,\ell,n)$ with $\ell > 0$ or $n > 0$ are stable as well. This is related to the question of what type of critical point (local minimum, local maximum or saddle point) of the energy functional they represent. The results of this article reveal the following properties. First, they indicate that all ground state configurations are stable under linearized perturbation modes of the spherically symmetric reduced system~(\ref{ellSis}) and that these configurations represent local minima of the total energy with respect to such perturbations. Second, they suggest that for $\ell > 0$ these minima are only local, i.e., they have a total energy which is larger than the corresponding energy of the ground state configurations with $\ell=0$ and the same value of $N$. Third, our results suggest that each excited configuration (i.e., each state with $n > 0$) is linearly unstable, possessing precisely $2n$ spherical linearized modes that grow exponentially in time. Fourth, they also indicate that each excited configuration represents a saddle critical point of the conserved energy functional.

Our stability results are consistent with previous studies on the linear stability of  $\ell$-boson stars with respect to spherical perturbations~\cite{Gleiser:1988ih, Gleiser:1988rq, Alcubierre:2021mvs} in the relativistic case. In these works, it is shown that the relativistic ground state configurations admit a stable branch which connects the Newtonian configurations with those of maximal mass. In particular, the linearized equations for mode solutions with time-dependency of the form $e^{-i\sigma t}$ are reduced to an eigenvalue problem of the form $\mathcal{H} v = \sigma^2 v$, where $\mathcal{H}$ is a two-channel Schrödinger operator. Since this operator is self-adjoint, $\sigma$ must be either real or purely imaginary, which is compatible with our findings for the ground state configurations. Interestingly, however, the method used in~\cite{Gleiser:1988ih, Gleiser:1988rq, Alcubierre:2021mvs} only works for the ground state solution since it requires the radial profile of the background scalar field to have a fixed sign. The results in our article indicate that excited states lead to the existence of imaginary eigenvalues; hence the underlying linear operator cannot be self-adjoint. Therefore, the results in the present article suggest that a liner stability analysis of the relativistic excited states requires a more general ansatz for the perturbed scalar field, which probably includes both factors $e^{-i\sigma t}$ and $e^{i\sigma^* t}$.

We end this article with a few comments regarding the physical implications of our results and a list of open questions. First, let us analyze the allowed range of numerical values for the total mass $M^{phys}$ and radius (which we define as the radius $R_{99}^{phys}$ of the centered ball containing $99\%$ of the mass) of the configurations $(N,\ell,n)$. These quantities scale with $N\mu$ and $1/(\mu^3 N)$, respectively, as can be seen from the definition of the dimensionless variables in Eq.~(\ref{AdVar}). Hence, for given $\ell$ and $n$ the object's mass and radius are determined by the two parameters $N$ and $\mu$. However, note that these parameters are not independent from each other. In order to be consistent with the nonrelativistic limit, $R_{99}^{phys}$ needs to be much larger than the Schwarzschild radius $R_s^{phys} := 2G M^{phys}/c^2$, which leads to the restriction\footnote{When $R_s^{phys}$ becomes comparable to $R_{99}$ one needs to consider the relativistic $\ell$-boson stars~\cite{Alcubierre:2018ahf} instead which have a maximum compactness corresponding to about half the Buchdahl limit for stable configurations~\cite{Alcubierre:2021psa}.}
\begin{align}
    \frac{R_s^{phys}}{R_{99}^{phys}}&=
\frac{(2N)^2}{(2\ell+1)R_{99}}\left(\frac{\mu}{m_{\text{pl}}}\right)^4\ll 1,
\label{PhysicRel}
\end{align}
where $m_{\text{pl}} = \sqrt{\hbar c/G} = 2.17643\times10^{-8}$kg is the Planck mass and $R_{99}$ is the dimensionless radius containing $0.99M$ of the dimensionless total mass for the correctly normalized radial profile. For example, for heavy masses of the order of the Higgs boson, such that $\mu\approx 1.25\times 10^{11} \text{eV}/c^2\approx 2.22\times10^{-25} \text{kg}$, the restriction~(\ref{PhysicRel}) yields $N\ll 10^{32}$ for values of $\ell\leq 10$ and assuming $R_{99} < 100$. This would give rise to objects with a maximal mass much lower than $10^{7}$kg (and configurations with such maximal masses would have a radius smaller than the Bohr one). However, particles with light masses are capable to fulfill the Newtonian restriction~(\ref{PhysicRel}) and have masses and radii compatible with typical astrophysical objects at the same time. For instance, objects formed of $N\approx10^{55}$ bosons of mass $\mu=10^{-3}\text{eV}/c^2\approx 1.78\times10^{-39} \text{kg}$ have masses and radii similar to a typical dwarf planet, for which $M^{phys}\approx 10^{16} \text{kg}$ and $R^{phys}\approx200\,\text{km}$. On the other hand, an ultralight mass $\mu=10^{-22}\text{eV}/c^2\approx 1.78\times10^{-58} \text{kg}$ with $N\approx10^{98}$ yields values compatible with dark matter galactic halos for which $M^{phys}\approx10^{10} \textup{M}_\odot$ and $R^{phys}\approx 1\,\text{Kpc}$. In both examples, the chosen number of fields $N$ fulfills the Newtonian restriction, which is $N\ll 10^{61}, 10^{99}$ for $\mu=10^{-3}, 10^{-22} \text{eV}/c^2$ respectively.

Second, let us comment on the timescales associated with the unstable modes of the unstable configurations $(N,\ell,n)$ with $n > 0$. Such configurations could still be considered to be stable for practical purposes if their lifetime is sufficiently large (e.g., of the order of the age of the Universe). For this reason, it is important to quantify these timescales. A referential value for them is defined by $t_{\text{life}} := 1/\lambda_R$, where $\lambda_R$ refers to the real part of the eigenvalue associated with the fastest growing mode. According to Eq.~(\ref{AdVar}) the physical lifetime scales like $1/(N^2\mu^5)$. For the configurations $(N,1,1)$, for which the fastest growing mode has $\lambda_R = 0.00619398$ and whose masses and radii correspond to the typical astrophysical objects discussed in the previous paragraph, one obtains the following values. For dwarf planets, one obtains short lifetimes of the order $t_{\text{life}}\approx 10^4 s\approx 2.8 \text{hr}$. For dark matter galactic halos the resulting lifetime is of the order $t_{\text{life}}=10^{13} s \approx 3.17\times 10^5$ yr, which is much shorter than the lifetime of a typical galaxy.

Of course, the numerical method used in this article has only been able to explore a finite parameter space; hence it would be interesting to put our stability results on a rigorous mathematical basis, and to prove that they are indeed true for arbitrary configurations $(N,\ell,n)$. However, a more pressing question is whether the ground state configurations $(N,\ell,0)$ with $\ell > 0$ are (linearly and nonlinearly) stable with respect to small time-dependent perturbations which are not necessarily spherical. This is related to the question of whether these configurations represent local minima or saddle critical points of the conserved energy functional with respect to arbitrary (i.e., not just spherical) variations. A further interesting problem consists in analyzing the stability properties of the nonrelativistic analogues of the multi-$\ell$ multi-state configurations found in~\cite{Alcubierre:2022rgp}, which include fields with different values of $\ell$ and $n$. We hope to address these questions in future work.

\subsection*{Acknowledgements}
It is a pleasure to thank Alberto Diez-Tejedor and Emilio Tejeda for  enlightening discussions and Fransisco S. Guzm\'an for reading the manuscript. This work was partially supported by CONACyT Network Projects No.~376127 ``Sombras, lentes y ondas gravitatorias generadas por objetos compactos astrofísicos'', by a CIC grant to Universidad Michoacana de San Nicolás de Hidalgo, and CONACyT-SNI. A.A.R. also acknowledges funding from a postdoctoral fellowship from ``Estancias Posdoctorales por México para la Formación y Consolidación de las y los Investigadores por México''. E.C.N. was supported by a CONACyT doctoral scholarship.

\appendix 

\section{Lagrangian formulation and total energy}
\label{App:LagrangeFormulation}

In this appendix, we provide a compact derivation of the relation between the total energy and the energy eigenvalues presented in Eq.~(\ref{Eq:FunctionalValueLBS}) which is based on a Lagrangian formulation for the SP system~(\ref{eqT}). To this purpose, we first introduce the column vector $\psi$, the row vector $\psi^*$, and the diagonal matrix $A$ as follows:
\begin{subequations}
\begin{eqnarray}
	\psi &=& (\psi_1, \psi_2, \ldots, \psi_J)^T,\\
	\psi^* &=& (\overline{\psi}_1, \overline{\psi}_2, \ldots, \overline{\psi}_J), \\
	A &=& \textbf{diag}(N_1, N_2, \ldots, N_J),
\end{eqnarray}
\end{subequations}
where we recall that $N_j$ ($j = 1, 2, \ldots, J$) refer to the number of particles in the state $\psi_j$. Here, the superscript $T$ denotes the transposed and $*$ the conjugate transposed. With this notation, the action corresponding to the SP system~(\ref{eqT}) can be written as~\cite{heisenberg1930physical}
\begin{equation}
    S[\psi,\psi^*,U] = \int_{\Omega} L(\psi,\psi^*,\partial_\mu \psi,\partial_\mu\psi^*,U,\partial_k U) d^4x,
\end{equation}
with an open subset $\Omega\subset \mathbb{R}^4$. Here and in the following, $(x^\mu) = (t,x^1,x^2,x^3) = (t,\vec{x})$, Greek and Latin indices run over $0,1,2,3$ and $1,2,3$, respectively, and Greek indices are raised and lowered by means of the Euclidean metric $\delta_{\mu\nu}$. The Lagrangian is given by 
\begin{align}
L =& -\frac{\hbar^2}{2\mu} \sum\limits_{k=1}^3(\partial_k\psi^{*}) A\partial_k \psi + \frac{i\hbar}{2}\left(\psi^*A\Dot{\psi} - \Dot{\psi}^*A\psi\right)\nonumber \\
    &- \mu U\psi^*A\psi - \frac{1}{8\pi G}\sum\limits_{k=1}^3 (\partial_k U)^2.
\label{Eq:Lagrangian}
\end{align}
It is simple to check that the corresponding Euler-Lagrange equations give rise to the SP system~(\ref{eqT}) and its complex conjugate.

According to Noether's theorem~\cite{Noether1918}, a continuous symmetry of the action gives rise to the conserved current
\begin{align}
\scalemath{0.91}{
J^\mu = \frac{\partial L}{\partial (\partial_\mu \psi)} \delta\psi + \delta\psi^* \frac{\partial L}{\partial (\partial_\mu \psi^*)}+\frac{\partial L}{\partial (\partial_\mu U)} \delta U
 + L\delta x^\mu,
\label{Current}
}
\end{align}
where $\delta\psi$, $\delta\psi^*$, $\delta U$, $\delta x^\mu$ refer to the first variations of the fields $\psi$, $\psi^*$, $U$ and the coordinates $x^\mu$ with respect to the action of the symmetry. For example, a translation gives rise to the conserved current
\begin{equation}
J^\mu = -\sum\limits_{\nu=0}^3 T^{\mu\nu}\delta x_\nu,
\end{equation}
with the stress energy-momentum tensor
\begin{align}
T^{\mu\nu} &= \frac{\partial L}{\partial (\partial_\mu \psi)} \partial^\nu \psi + \partial^\nu\psi^* \frac{\partial L}{\partial (\partial_\mu \psi^*)}+\frac{\partial L}{\partial (\partial_\mu U)} \partial^\nu U 
\nonumber\\
 & - L\delta^{\mu\nu}.
\label{Eq:Tmunu}
\end{align}
In particular, assuming that the fields decay sufficiently fast at infinity, it follows that the total energy
\begin{equation}
\int\limits_{\mathbb{R}^3} T^{00} d^3 x
\end{equation}
is conserved in time. Using Eqs.~(\ref{Eq:Lagrangian},~\ref{Eq:Tmunu}), one can check that this energy coincides precisely with the conserved energy functional $\mathcal{E}$ in Eq.~(\ref{ecFuncGenReduced}).

For the following, we consider the scale transformation (cf. Eq.~(\ref{AdVar}))
\begin{align}
    \begin{array}{rcl}
    t\mapsto \tilde{t} = \Lambda^{-2} t,&\qquad& x^i \mapsto \tilde{x^{i}} = \Lambda^{-1} x^i,\\[0.2cm]
    \psi\mapsto \tilde{\psi} = \Lambda^2 \psi, &\qquad& \psi^* \mapsto \tilde{\psi^*} = \Lambda^2 \psi^*,\\[0.2cm]
     U \mapsto \tilde{U} = \Lambda^2 U, & &
    \end{array}
\end{align}
with an arbitrary positive factor $\Lambda > 0$. This transformation implies that the Lagrangian rescales according to $\tilde{L} := L(\tilde{\psi},\tilde{\psi}^*, \partial_\mu\tilde{\psi},\partial_\mu\tilde{\psi}^*,\tilde{U},\partial_i\tilde{U}) = \Lambda^6 L(\psi,\psi^*,\partial_\mu\psi,\partial_\mu\psi^*,U,\partial_i U)$; hence it leaves the equations of motion invariant. The action, however, is not invariant but satisfies the relation
\begin{align}
\int_{\tilde{\Omega}} \tilde{L} 
d^4\tilde{x} = \Lambda\int_{\Omega}  L d^4 x,
\end{align}
which, upon variation (i.e., derivation with respect to $\Lambda$ evaluated at $\Lambda = 1$), yields the following relation for solutions of the Euler-Lagrange equations:
\begin{align}
\sum\limits_{\mu=0}^3\partial_\mu J^\mu = L.
\end{align}
Although the current $J^\mu$ is not conserved in this case, one still obtains a useful relation by integrating this equation over $\mathbb{R}^3$. Assuming sufficiently rapid decay of the fields at infinity, one obtains
\begin{align}
\frac{d}{dt} \int\limits_{\Omega} J^0 d^3 x = \int\limits_{\Omega} L d^3 x, \label{Charge}
\end{align}
where, using Eq. (\ref{Current}), one finds the following expression for $J^0$:
\begin{align}
J^0 = 2tT^{00} + \sum\limits_{k=1}^3 T^{0k}x_k. \label{Current0}
\end{align}
For the particular case of time-harmonic solutions of the form
\begin{align}
\psi_j(t,\vec{x}) = e^{\sfrac{-i E_j t}{\hbar}} u_j(\vec{x}),\qquad j=1,2,\ldots,N_J,
\end{align}
with functions $u_j$ which are independent of $t$ and satisfy the normalization condition $(u_j,u_k) = \delta_{jk}$, one obtains from Eqs.~(\ref{Eq:Lagrangian},~\ref{Charge},~\ref{Current0}) the relation
\begin{align}
3\mathcal{E}[u] = \sum\limits_{j = 1}^{J} N_j E_j. \label{GeneralRelat}
\end{align}
In particular, for the $\ell$-boson stars considered in this article this yields\footnote{Note that in this appendix we work in physical units such that $\mathcal{E} = \mathcal{E}^{phys}$ and $E_\ell = E_\ell^{phys}$.}
\begin{equation}
\mathcal{E}[u] = \frac{N}{3} E_\ell = \frac{(2\ell + 1)K}{3} E_{\ell},
\end{equation}
which proves the relation~(\ref{Eq:FunctionalValueLBS}).

\section{First and second variations of the energy functional}
\label{ApendVf}

In this appendix, we compute the first and second variations of the reduced energy functional $\mathcal{E}_\ell$ defined in Eq.~(\ref{ecFuncellBS}), which reads
\begin{align}
    \mathcal{E}_{\ell}[f] &= \int_0^\infty \left[ \abs{\partial_r f(r)}^2+\frac{\ell(\ell+1)\abs{f(r)}^2}{r^2}\right] r^2 dr\nonumber\\
    &-\frac{1}{2} \int_0^\infty\int_0^\infty\frac{\abs{f(r)}^2 \abs{f(\tilde{r}))}^2}{r_{>}} r^2 \tilde{r}^2 dr d\tilde{r},
\label{Ap1Eq2}
\end{align}  
where we recall the notation $r_{>}=\max  \left\lbrace r, \tilde{r} \right\rbrace$. We also prove the relation~(\ref{Eq:FundamentalRelation}) between the second variation of $\mathcal{E}_\ell$ and the expectation value of the operator $\hat{Q}$ defined in Eq.~(\ref{EqOpe}).

To perform the variation, we expand the wave function $f$ in the following form:
\begin{align}
\scalemath{0.965}{
f(t, r)=f^{(0)}(t, r)+\epsilon \delta f(t, r) +\frac{\epsilon^2}{2}\delta^2 f(t, r) + \mathcal{O}(\epsilon^3),
\label{Ap1Eq3}
}
\end{align}
where $f^{(0)}(t, r)$ denotes the (real-valued) background field and $\delta f(t, r), \delta^2 f(t, r)$ denote their first and second order (complex-valued) perturbations, respectively. The $n$th variation of $\mathcal{E}_\ell$ is defined as
\begin{align}   &\delta^n\mathcal{E}_{\ell}
 :=\frac{d^n}{d\epsilon^n}\eval{\mathcal{E}_{\ell}[f]}_{\epsilon=0}.\label{Ap1Eq4}
\end{align}
After some manipulations a straightforward calculation yields
\begin{subequations}\label{Ap1Eq56}
\begin{align}
   &\delta\mathcal{E}_{\ell}=2\Re \int_0^\infty dr\, r^2 \, \delta f^{*} \hat{\mathcal{H}^{(0)}}_\ell f^{(0)},
   \label{Ap1Eq5}\\
    &\delta^2\mathcal{E}_{\ell}=\Re \int_0^\infty dr\, r^2 \bigg( \delta^2 f^{*} \hat{\mathcal{H}}^{(0)}_\ell f^{(0)} + \delta f^{*}\hat{\mathcal{H}}^{(0)}_\ell\delta f\bigg)
    \nonumber\\
    & -2\int_0^\infty\int_0^\infty dr d\tilde{r}\, r^2\tilde{r}^2  \frac{\Re\big(f^{(0)} \delta f^{*}\big)\Re\big(\tilde{f}^{(0)} \delta \tilde{f}\big)}{r_{>}},\label{Ap1Eq6}
\end{align}
\end{subequations}
where we have used the definition of the operator $\hat{\mathcal{H}}^{(0)}_\ell$ defined in Eq.~(\ref{Eq:Hell0}) and where, for notational simplicity, we have omitted the argument $r$ of the functions, using the {\it tilde} to indicate that the function is evaluated at $\tilde{r}$ instead of $r$.

For the particular case that $f^{(0)}$ is a solution of the system~(\ref{ellSis1}) with a harmonic temporal dependence as in Eq.~(\ref{Hansatz}) we have $\hat{\mathcal{H}}^{(0)}_\ell f^{(0)} = E_\ell f^{(0)}$ and obtain
\begin{subequations}\label{Ap1Eq78}
\begin{align}
\delta\mathcal{E}_{\ell}&=
2 E_{\ell}\Re\left(\delta f, f^{(0)}\right)_{L^2},
\label{Ap1Eq7}\\
\delta^2\mathcal{E}_{\ell}&=\left(\delta f, \hat{\mathcal{H}}^{(0)}_\ell \delta f \right)_{L^2} 
+ E_\ell\Re \left(\delta^2 f, f^{(0)} \right)_{L^2}
\nonumber\\
& -2\int_0^\infty\int_0^\infty dr d\tilde{r}\, r^2\tilde{r}^2  \frac{\Re{f^{(0)} \delta f^{*}}\Re{\tilde{f}^{(0)} \delta \tilde{f}}}{r_{>}}, \label{Ap1Eq8}
\end{align}
\end{subequations}
where we recall that $(\cdot,\cdot)_{L^2}$ refers to the standard scalar product defined in Eq.~(\ref{innerProd}). Taking into account the normalization condition Eq.~(\ref{EqNorma}), which implies that $(f,f)_{L^2}$ is constant, such that
\begin{align}
\Re\left(\delta f, f^{(0)} \right)_{L^2} =0,\quad
\Re\left(\delta^2 f, f^{(0)} \right)_{L^2}  =
-\left(\delta f, \delta f \right)_{L^2},
\end{align}
we conclude that $\delta\mathcal{E}_{\ell} = 0$ and
\begin{align}
\delta^2\mathcal{E}_{\ell}&=\left(\delta f, [\hat{\mathcal{H}}^{(0)}_\ell-E_{\ell}] \delta f \right)_{L^2}\nonumber\\
& -2\int_0^\infty\int_0^\infty dr d\tilde{r}\, r^2\tilde{r}^2  \frac{\Re{f^{(0)} \delta f^{*}}\Re{\tilde{f}^{(0)} \delta \tilde{f}}}{r_{>}}.
\label{Ap1Eq9}
\end{align}
Comparing this expression with the definitions~(\ref{EqOpe}, \ref{Eqiner}) of the operator $\hat{Q}$ and the inner product $\braket{\cdot}{\cdot}$ we arrive at the fundamental relation between the second variation of $\mathcal{E}_\ell$ and the expectation value of $\hat{Q}$:
\begin{align}
   \braket{A}{\hat{Q} A} = \delta^2\mathcal{E}_{\ell}[A_R] + \delta^2\mathcal{E}_{\ell}[A_I],
\end{align}
where we have set $A = A_R + i A_I$ and the notation $\delta^2\mathcal{E}_{\ell}[A_i]$ refers to the second variation evaluated at \mbox{$\delta f = A_i$}. This relation allows one to connect the sign of the second variation of $\mathcal{E}_\ell$ with the one of the expectation values of $\hat{Q}$ and plays a crucial role in our stability analysis.

\section{Determination of the energy eigenvalues}
\label{ApendEE}

In this appendix we present the methodology to compute the energy eigenvalue $E_{\ell}$ corresponding to a solution $(\sigma_\ell^{(0)},u^{(0)})$ of the background system (\ref{Ec2.2.2.1}). Our procedure is a straightforward generalization to $\ell\geq 0$ of the prescription given in~\cite{Moroz:1998dh}.

Recall the expression for the gravitational potential
\begin{equation}
U^{(0)}(t, r)=\triangle^{-1}_{s}\left(\abs{\sigma_\ell^{(0)}}^2\right)=E_{\ell}-u^{(0)}(r),
\end{equation}
which is defined in terms of the shifted potential $u^{(0)}$ defined in subsection~\ref{IIIA}. Since $U^{(0)}$ vanishes at infinity, we can in principle calculate the energy eigenvalues $E_\ell$ by taking the asymptotic limit
\begin{equation}
E_{\ell}=\lim\limits_{r\to\infty} u^{(0)}(r).\label{ApBeq1}
\end{equation}
However, the problem is that with the shooting method used in this article, the asymptotic value of $u^{(0)}$ is out of reach. To deal with this problem we use the following approximation.

Recalling the relation $\sigma^{(0)}_{\ell}(r)=r^{\ell}\sigma(r)$, integrating the system~(\ref{Ec2.2.2.1}) twice with respect to $r$, using integration by parts, and taking into account the boundary conditions~(\ref{BCOr0}), we arrive to the equivalent integral system,
\begin{subequations}
\begin{align}
\sigma(r)&=\sigma_0+\int_{0}^{r}\frac{u^{(0)}(x)\sigma(x)}{2\ell+1} \left[\left(\frac{x}{r}\right)^{2\ell+1}-1\right]x dx,\\
u^{(0)}(r)&=u_0+\int_{0}^{r}\abs{\sigma(x)}^2\left(\frac{x}{r}-1\right)x^{2\ell+1} dx.\label{IntPot}
\end{align}
\end{subequations}
Differentiating Eq.~(\ref{IntPot}) with respect to $r$ yields
\begin{equation}\label{du0}
    \frac{du^{(0)}(r)}{dr}=-\frac{1}{r^2}\int_{0}^{r}\abs{\sigma(x)}^2 x^{2(\ell+1)} dx,
\end{equation}
which implies that $u^{(0)}$ is monotonically decreasing. Consequently, the gravitational potential $U^{(0)}$ is monotonically increasing to zero (which is consistent with the behavior shown in Fig.~\ref{Upplot}). Further, since $\sigma(x)$ is exponentially decaying as $x\to \infty$, one can expand $u^{(0)}$ in powers of $r^{-1}$,
\begin{equation}
u^{(0)}(r) = E_\ell +\frac{M}{r}+O(r^{-3}),\label{UpowerS}
\end{equation}
where the constants $E_\ell$ and $M$ are given by
\begin{subequations}\label{EqC6}
\begin{align}
E_\ell &=u_{0}-\int_{0}^{\infty} x^{2\ell+1}\abs{\sigma(x)}^2 dx,\label{ApB5a}\\
M &=\int_{0}^{\infty} x^{2(\ell+1)}\abs{\sigma(x)}^2 dx.
\end{align}
\end{subequations}
Note that $M$ is the integral over the mass density $|\sigma_\ell^{(0)}(r)|^2$ times $r^2$; hence it represents the dimensionless total mass of the configuration.

The relation~(\ref{ApB5a}) provides an alternative method for computing $E_\ell$, provided $u_0$ and the radial profile of $\sigma(r)$ are known.\footnote{In practice, the integrals in Eq.~(\ref{EqC6}) cannot be computed over the whole range $0 < x < \infty$ since the profile $\sigma$ obtained from the shooting algorithm is only known up to some maximum radius $r_{\text{max}}$. However, due to the exponential decay of $\sigma$, one can truncate the integral at $r_{\text{max}}$; the contributions from the interval $r_{\text{max}}<x<\infty$ do not affect the results at the level of the significant figures reported in this work.
}
However, recall that the numerical solution obtained from the shooting method does not directly satisfy the normalization condition Eq.~(\ref{EqNorma}), which means that Eq.~(\ref{ApB5a}) yields the unrescaled energy eigenvalue. There are two options to compute the correctly scaled eigenvalue. The first one consists in rescaling the solution $(\sigma_\ell^{(0)},u^{(0)})$  using the relations in Eqs.~(\ref{reescaling}) with $\Lambda=(2\ell+1)/M$ (see Eq.~(\ref{LambConst})) and then compute $E_{\ell}$ according to Eq.~(\ref{ApB5a}). The second option is to first compute the unrescaled value of $E_\ell$ using Eq.~(\ref{ApB5a}) and then use the transformation $E^\Lambda_{\ell}=\Lambda^2 E_{\ell}$, which yields
\begin{equation}
E_{\ell}^{\Lambda}=\frac{(2\ell+1)^2}{M^2}E_{\ell}.
\label{Eq:ELambdaell}
\end{equation}

The results reported in this article are based on the second option which determines $E_\ell$ from Eq.~(\ref{ApB5a}) from the unrescaled profiles. The (correctly rescaled) dimensional eigenvalue corresponding to $N=K(2\ell+1)$ particles is obtained from the formula
\begin{equation}
E^{phys}_{\ell}=2\mu v_c^2 K^2 E^{\Lambda}_{\ell},
\end{equation}
with $E^{\Lambda}_{\ell}$ given by Eq.~(\ref{Eq:ELambdaell}). However, we have also checked the results using the first option.

To close this appendix, we notice that the physical mass is given by
\begin{align}
M^{phys} =  K \mu \Lambda M.
\end{align}
For configurations with radial profiles $\sigma^{\Lambda}$ satisfying the normalization conditions Eq.~(\ref{EqNorma}) one has $M = (2\ell+1))/\Lambda$, such that the physical mass is
\begin{equation}
M^{phys} = (2\ell+1) K \mu = N\mu,
\end{equation}
as expected.

\section{Convergence and independent residual analysis}
\label{apend2}

\begin{figure*}
	\centering	
	\includegraphics[width=18.cm]{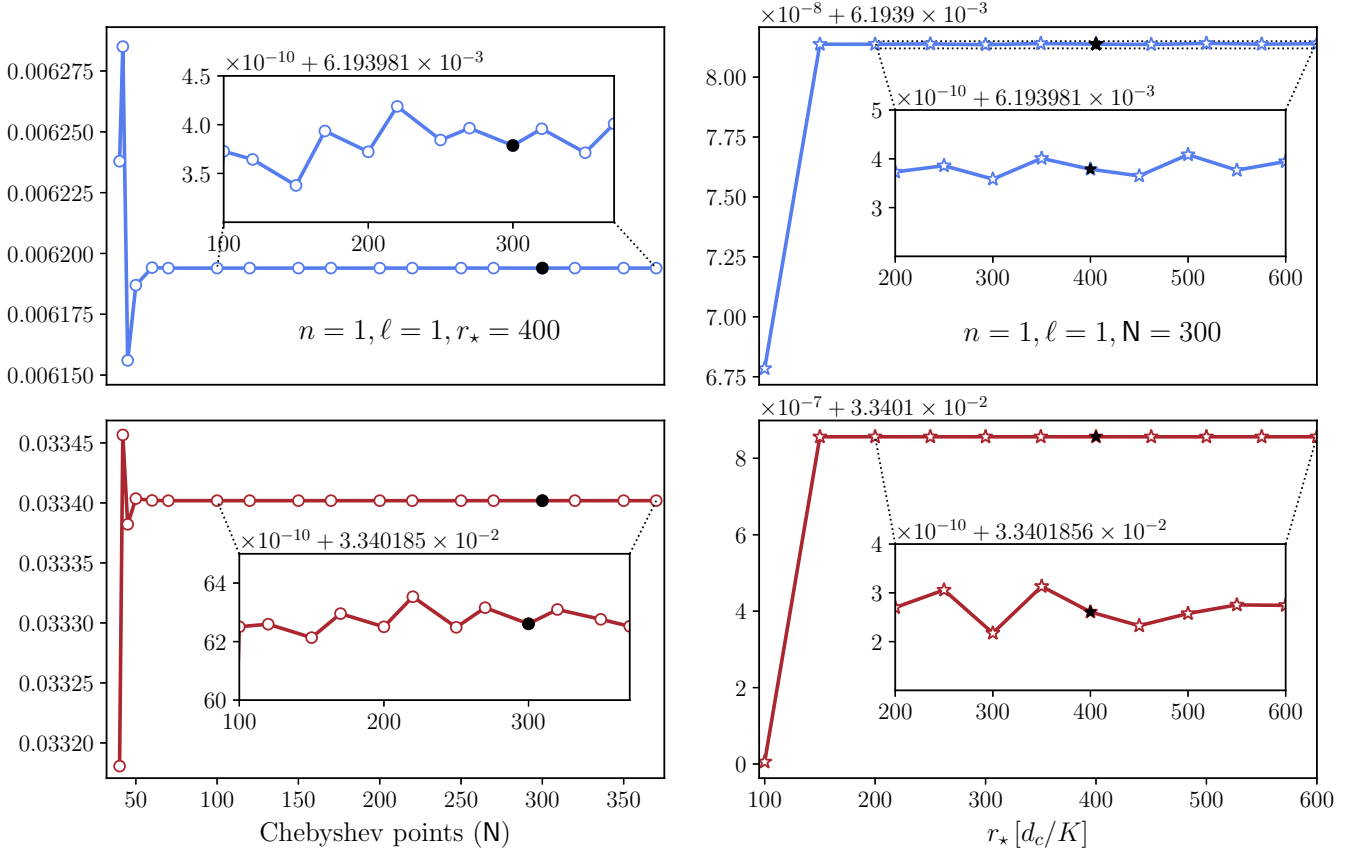}
	\caption{Convergence study for the eigenvalue $\lambda = 0.00619398 + 0.03340186i$ corresponding to the background configuration with $(n,\ell)=(1,1)$. The plots in the left column show the numerical values for $\lambda$ for fixed $r_\star = 400$ and varying number of Chebyshev points $\mathsf{N}$. The plots in the right column show the same quantity, now for fixed value of $\mathsf{N} = 300$ and varying $r_\star$. Here, the top and bottom plots show the real ($\lambda_R$) and imaginary ($\lambda_I$) parts of $\lambda$. The points corresponding to the dark circle and star correspond to the values reported in this article, and as can be seen, they lie in the region in which the relative error is comparable or smaller than $10^{-7}$.}\label{ApEst}
\end{figure*}

\begin{figure*}
	\centering	
	\includegraphics[width=18.cm]{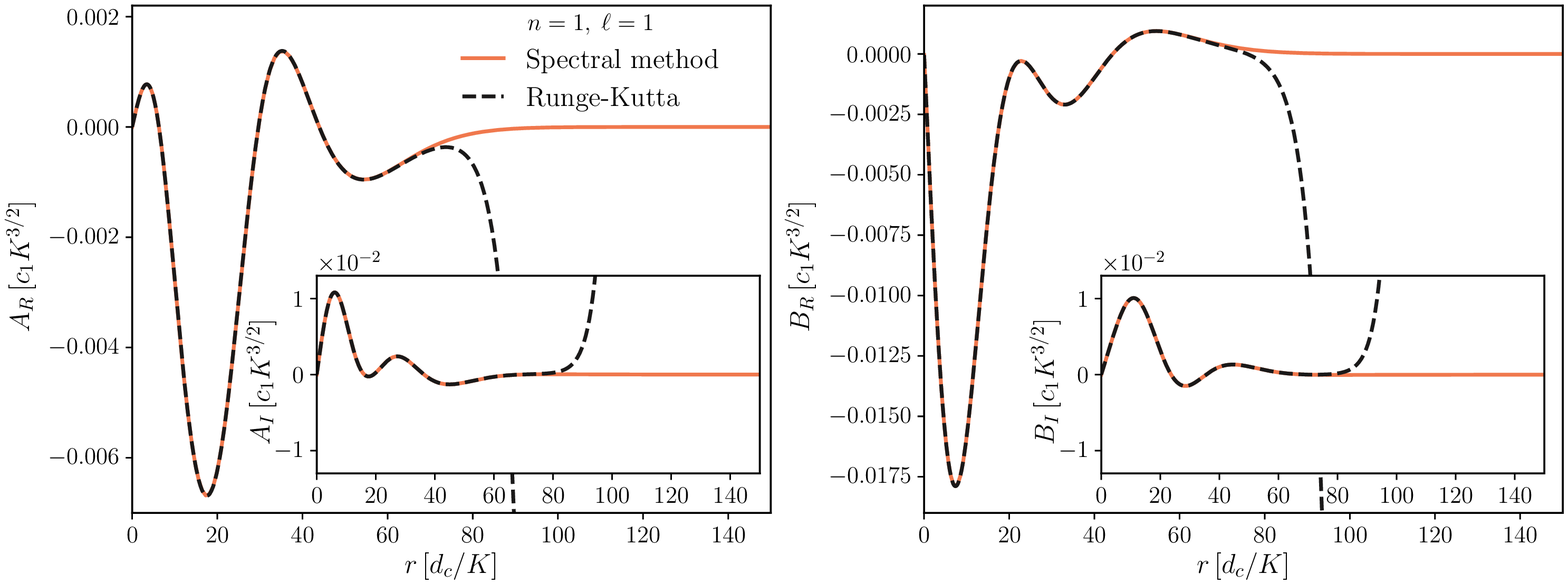}
	\caption{Comparison between the linearized mode $(A, B)$ obtained from the spectral and Runge-Kutta methods for the background configuration corresponding to $(n, \ell)=(1, 1)$ and the eigenvalue $\lambda = 0.00619398 + 0.03340186i$. As can be appreciated from this plot, the Runge-Kutta results are consistent with the spectral ones up to a radius of the order $r\sim70$.\label{checkP}}
\end{figure*}

In this appendix we check the convergence of our pseudo-spectral collocation method used to compute the eigenvalues of the linearized system~(\ref{Ec2.2.2.2}). Furthermore, we present an independent residual analysis based on a Runge-Kutta integration of the linearized equation.

As discussed in subsection~\ref{IVA} our method involves mapping the truncated physical domain $[0, r_{\star}]$ on the computation domain $x\in[-1,1]$ by means of the transformation
\begin{align}
r = r_{\star}\left(\frac{x+1}{2}\right). \label{Map1}
\end{align}
Alternatively, we consider the transformation
\begin{equation}
r = 2S\left( \frac{1}{1-y} - \frac{1}{2}\right), \label{Map2}
\end{equation}
with $S > 0$ a characteristic distance, which maps the whole physical domain $[0,\infty]$ onto $y\in [-1,1]$ and is independent of $r_\star$. Recall that we discretize the computational domain using a Chebyshev distribution with $\mathsf{N}+1$ points, which yields the best accuracy for homogeneous Dirichlet boundary conditions (see Chapter-V in~\cite{trefethen2000spectral}). Therefore, the numerical error depends on the two parameters $r_\star$ and $\mathsf{N}$ (in the case of the compactified $y$-domain the parameters are $S$ and $\mathsf{N}$), and it is necessary to analyze the convergence of the numerical results with respect to these parameters. 

The results shown in this paper are computed using the map~(\ref{Map1}), whereas the alternative map~(\ref{Map2}) is used to validate them. The ideal choice for $\mathsf{N}$ and $r_\star$ depends on the background solution; for instance, as can be seen from Fig.~\ref{Wfplot}, a large node number leads to more extended configurations which require higher values of $\mathsf{N}$ and $r_\star$ than the ground states to achieve the same accuracy (see e.g., Program~$15$ in~\cite{trefethen2000spectral} for an illustrative example). We found that the empirical choices $r_{\star}:=200(n+1)$, with $n$ the background solution's node number, and $\mathsf{N}:=3r_\star/4$ lead to acceptable results (in particular, it guarantees an accuracy of seven significant digits for the first eigenvalues).

Figure~\ref{ApEst} shows a convergence study for the eigenvalue with nonzero real part, $\lambda = 0.00619398 + 0.03340186i$, corresponding to the configuration with $(\ell,n)=(1, 1)$ (cf. Table~\ref{tabla2} and Fig.~\ref{Efplot} for the associated eigenfunction), in which both parameter values $\mathsf{N}$ and $r_\star$ are varied. This study indicates that our choice yields relative errors comparable or smaller than $10^{-7}$ also for this configuration. As is also visible from these plots, it is possible to choose $\mathsf{N}$ much less than $3r_\star/4$ keeping a comparable accuracy for the eigenvalue $\lambda$. However, in this case, we have found that the zero eigenvalue (corresponding to the zero mode discussed in the subsection~\ref{IIE}) may not be zero to machine precision anymore and may be confused with a non-zero eigenvalue.

After this convergence study, we turn our attention to the independent residual analysis. To this purpose, we implemented an explicit 5(4)-order Runge-Kutta routine and integrate the system~(\ref{Ec2.2.2.2}) from the origin $r=0$ outwards, fixing the eigenvalue found from the spectral analysis. Similar to the treatment of the background equations, we rescale the perturbed fields $(a,b)$ according to $(a,b) = r^{\ell+1}(X_2, X_3)$, such that the new fields $(X_2, X_3)$ are regular at $r=0$ (see  subsection~\ref{IVA}). Further, we introduce the new field
\begin{align}
X_1(r) := \frac{2}{r}\left(\frac{d^2}{dr^2} \right)^{-1}\left[\sigma_\ell^{(0)} a \right],
\label{Eq:X1}
\end{align}
with $\left(\frac{d^2}{dr^2} \right)^{-1}$ denoting the inverse of the second-derivative operator with homogeneous Dirichlet conditions at $r=0$ and $r=\infty$. With this notation, the system~(\ref{Ec2.2.2.2}) can be written as the following first-order system of ordinary differential equations
\begin{subequations}\label{RKfirstO}
\begin{align}
    X_{1}' = & Y_1,\\
    X_{2}' = & Y_2,\\
    X_{3}' = & Y_3,\\
    Y_{1}'=& 2 r^{\ell} X_2 \sigma_{\ell}^{(0)}-\frac{2Y_{1}}{r},\\
    Y_{2}'=& -i\lambda X_3-X_2 u^{(0)}+\frac{X_1 \sigma_{\ell}^{(0)}}{r^{\ell}}-\frac{2(\ell+1)Y_{2}}{r},\\
    Y_{3}'=& -i\lambda X_2-X_3 u^{(0)}-\frac{2(\ell+1)Y_{3}}{r},
\end{align}
\end{subequations}
with $(\sigma_{\ell}^{(0)}, u^{(0)})$ the background fields and $\lambda$ an eigenvalue corresponding to an associated linear mode. The system is numerically solved subject to the boundary conditions
\begin{subequations}\label{Eq:xi}
\begin{align}
    X_1(0)=&x_1,\quad X_2(0)=x_2,\quad X_3=x_3,\\
    Y_1(0)=&0,\quad Y_2(0)=0,\quad Y_3(0)=0,
\end{align}
\end{subequations}
where here the conditions $Y_i(0) = 0$ follow from a standard regularity requirement on the fields and the values $x_i$ are computed from the respective fields $(a,b)$ obtained from the spectral method. In practice, we specify the data at the first grid point after the origin in order to avoid the singular $1/r$ terms, and we use Eq.~(\ref{Eq:X1}) to determine $x_1$.

Figure~\ref{checkP} shows the corresponding results for the same background configuration $(n,\ell)=(1,1)$ and eigenvalue $\lambda = 0.00619398 + 0.03340186i$ as in the convergence study. Shown are the fields $(A, B) = r^\ell(X_2, X_3)$ computed from the Runge-Kutta method described here and the same fields obtained from the spectral calculation described in subsection~\ref{IVA}. Despite the sensitive dependency of the Runge-Kutta solution on the data at $r=0$ (i.e., the values of $\lambda$ and $x_i$ in Eq.~(\ref{Eq:xi})), we see from this figure that both results are consistent at least up to radii $r\sim 70$. 

\bibliographystyle{unsrt}
\bibliography{ref.bib} 

\end{document}